# Contextual Regression: An Accurate and Conveniently Interpretable Nonlinear Model for Mining Discovery from Scientific Data

Chengyu Liu[1], Wei Wang[1,2]

**Abstract**

Machine learning algorithms such as linear regression, SVM and neural network have played an increasingly important role in the process of scientific discovery. However, none of them is both interpretable and accurate on nonlinear datasets. Here we present contextual regression, a method that joins these two desirable properties together using a hybrid architecture of neural network embedding and dot product layer. We demonstrate its high prediction accuracy and sensitivity through the task of predictive feature selection on a simulated dataset and the application of predicting open chromatin sites in the human genome. On the simulated data, our method achieved high fidelity recovery of feature contributions under random noise levels up to ±200%. On the open chromatin dataset, the application of our method not only outperformed the state of the art method in terms of accuracy, but also unveiled two previously unfound open chromatin related histone marks. Our method can fill the blank of accurate and interpretable nonlinear modeling in scientific data mining tasks.

## Introduction

Predictive models are important tools for data analysis. Linear models, such as linear[1] and logistic regression[2], have been essential prediction methods for a very long time. They can provide adequate prediction accuracy while their linear formulation makes them suitable for inferring relationship between prediction target and features. Recently, complex nonlinear models have gained increasing popularity as an alternative to linear models. Methods such as kernel SVM[3], neural network[4] and decision trees[5] can achieve better prediction performance than the linear models especially on nonlinear datasets. In biology, for instance, deep convolutional neural network (DCNN) methods such as DeepBind[6] and DeepSea[7] have been developed for scanning the genome for regions of interest with state of the art accuracy.

This accuracy improvement, however, also comes with a cost: The parameters of a nonlinear model are hardly human interpretable. This not only makes model improving and debugging difficult but also prevents us from acquiring knowledge from these models and validating their findings, which is particularly important when controversial conclusions can be reached from the prediction results[8,9]. While examining the filters[10,11] in a DCNN can offer some insight of the important feature combinations, it cannot provide quantification of the feature contributions: in the later layers, the features are processed through multiple rounds of matrix multiplication, addition and neuron activation which makes their contributions to the output intractable. Besides, this approach cannot be applied to other neural networks or machine learning methods such as feedforward neural network (FNN)[12] or Long Short Term Memory (LSTM)[13] neural network. To fill this gap, many well-thought and intriguing methods such as DeepLift[14], LIME[15] and SHAP[16] have been developed for quantifying feature contributions in general machine

[1] Department of Chemistry and Biochemistry, University of California San Diego, La Jolla, CA 92093-0359
[2] Department of Cellular and Molecular Medicine, University of California San Diego, La Jolla, CA 92093-0359

learning models. All of them are based on the same overall strategy. When using these methods, the user first needs to choose a reference data point with the target value of interest, a genomic region with enhancer label of value 1 for example. Then the program will generate permuted data points by making small changes to the input features, which forms an interrogating dataset. The interrogating dataset are fed into a trained model of user's interest and the output of the model are collected. From the magnitude of the output change caused by the input change, an interpreting model that describes the feature contribution to the prediction result can be generated for that data point. This approach, however, has a few practical problems. First, since interpretation is based on each user selected reference data point, it is inherently a "local" model and thus can be overfitted to that point. Second, the training and interpretation processes are decoupled, which blocks the possibility of real time model monitoring during training. Third, the user is required to be familiar with the dataset in order to pick representative reference points, which is not an easy task particularly in scientific data when (i) the prediction target value is a continuous value other than binary; (ii) the dataset contains natural noise that causes the target value to deviate from its real value; (iii) the data are of large quantity and diversity.

We have developed a new method called contextual regression to address these challenges. It can quantify feature contributions during training without user intervention while preserving the accuracy achieved by a complex nonlinear model: we train the model to learn an embedding function to map each feature vector to a corresponding linear model that can predict the target value most accurately (Figure 1). In principle, values of the elements in the input feature vector describe its context and the embedding serves as a classifier of the context. It generates a continuous value vector as the class of the context which is similar to the mechanism of attention[17] and word2vec[18]. We call this class of the context "context weight" and thus this method the "contextual regression." In this way, the contribution of each feature can be inferred from the statistics of the context weights on the dataset. We demonstrated its accuracy and high interpretability through quantifying feature contributions to prediction accuracy in a simulated dataset with known "ground truth" and an application to the open chromatin prediction.

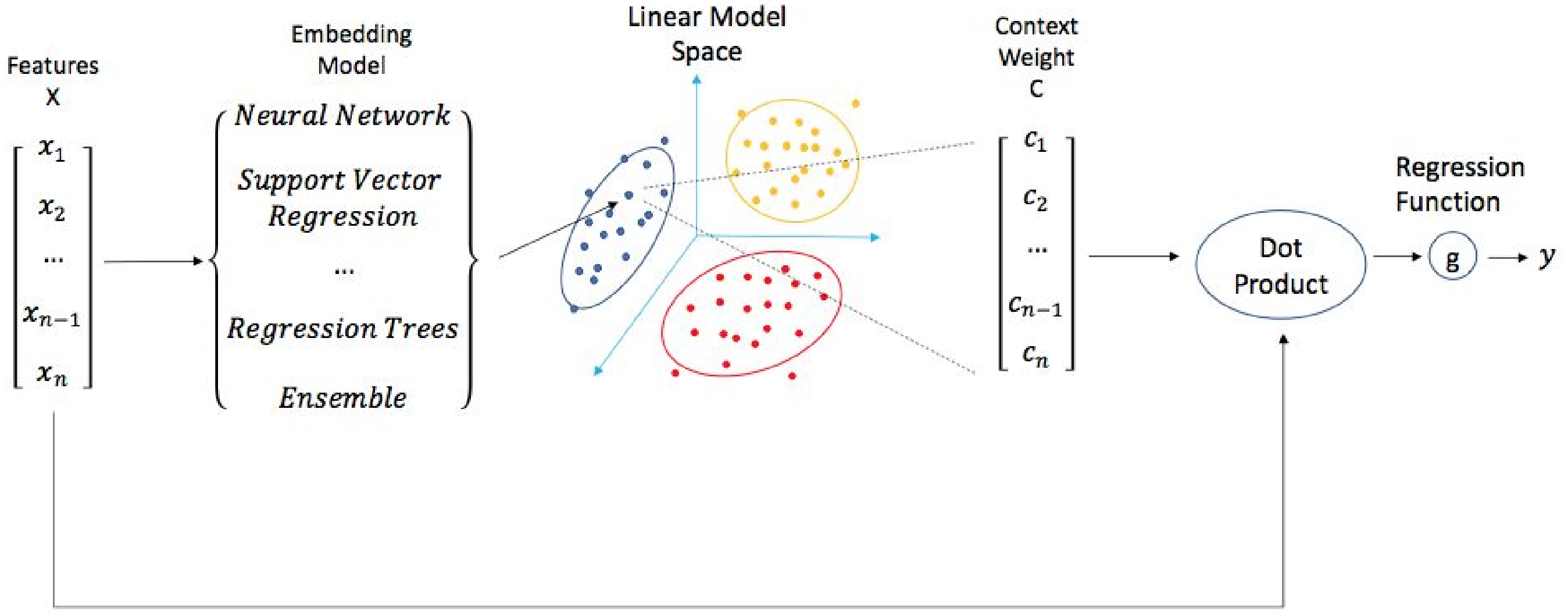

**Figure 1.** A graphic demonstration of the contextual regression model. Dot product apply the linear model, which is the context weight vector C, to the feature vector X. The embedding model can be any machine learning methods that can produce a vector output from a vector input. Different colors of points in the linear model space is an illustration that they can be classified into different subtypes.

## Results

### The contextual regression method

As shown in Figure 1, our model aims to learn a pseudo-linear model that approximates the function:

$$y = c(x_1, x_2, x_3, ..., x_n) \approx g(c(x) \cdot x + b) = g(\sum_{i=1}^{n} c_i(x_1, x_2, x_3, ..., x_n)x_i + b)$$

where y is the prediction target, $x_i$ are the features, b is a constant, g is the regression function (for example, it is an identity function for linear regression and is a logistic function for logistic regression), and c is the embedding function, which calculates the context weight from the feature vector x as the linear model for the prediction of y at that data point.

Thus, rather than trying to fit all the data with a single linear model, our method learns many linear models and apply different ones at different data points according to their values. Like illustrated in Figure 1, the embedding network takes a data point (represented as a vector) as its input, and outputs a linear model (the context weight, represented as a vector). Then the data point is dot-producted with the linear model to output the prediction. This method, in its property, can handle nonlinear relationship to make accurate predictions while still producing human-readable linear relationship between features and the prediction target. One potential caveat, an old problem in predictive modeling, is the existence of alternative model on the dataset, i.e., very different model parameters can yield similar prediction accuracy. Since a descent direction for y is not necessarily a descent direction for each individual $c_i$ (see supplement for detail), this scenario can possibly happen. We have developed the utopia penalty technique to address it. This technique reveals equivalent or interchangeable features by adding a term to the cost function that penalizes unbalanced context weights. The setup and an application example of it can be found in the supplement.

We evaluated the performance of the proposed method using a simulated dataset and a DNaseI hypersensitivity (DHS) dataset[19]. In both cases, the features have distal relationship and thus we used the bidirectional-Long Short Term Memory (LSTM) neural network, a specialized neural network model for distally related data, to be the main component of our embedding function.

### Evaluating the Contextual Regression model on simulated data

The simulated dataset includes an artificial “ground truth” in its feature-target relationship that we wish to find. We sampled features $x_i$ from an exponential distribution to (1) reduce the possibility of alternative models such that we can test the rule extraction ability of our method, and (2) imitate the scenario that the feature values differ in order of magnitude such as the sequencing read count in biological data. In the simulated data, we consider 50 features $x_i$ that are in sequential order. Each element $c_i$ of the context weight vector C is determined by the value of $x_i$ and its neighbors. The “ground truth” context weights $c_i$ and then the target value y are calculated from $x_i$ using the formula below. The context weight formula is chosen such that it is composed of common functions observed in natural data (linear, square and square root) and includes relationship with the neighboring elements. More details about the setup can be found in the supplement.

$$y = \sum_{i=1}^{n} c_i(x_1, x_2, x_3, ..., x_n)x_i \quad (1)$$

$$c_i(x_1, x_2, x_3, ..., x_n) = \frac{1}{1000}[\sum_{j=i-1}^{1} 0.3^{i-j}u(x_j) + \sum_{k=i+1}^{n} 0.3^{k-i}u(x_k) + x_i] \quad (2)$$

$$u(x_k) = \sqrt{500x_k} + \frac{x_k^2}{500} \quad (3)$$

Our model was tested under 5 noise levels of ±0% to ±80%. When adding the noise, a random number inside the noise range was sampled uniformly (for instance, between -80% to +80% for noise level ±80%) and this fraction was added or subtracted from the target value y. The model was trained on 70% of the randomly selected data (training set) and then its fitness was tested on the remaining 30% data (testing set). We used root mean square error (RMSE) as the measure of prediction accuracy. Cosine distance measures the angle difference between two vectors, and thus is a good measure of vector similarity. So for the measure of rule extraction fidelity, we used root mean square cosine distance (RMSCD), the cosine distance version of RMSE, between the "ground truth" weights and the weights produced by our embedding network at their corresponding data points. Since the target values contain noise, it is impossible for the model to achieve 100% accuracy. So we compared RMSE with the expected error under the corresponding noise level, which was calculated by integrating the percent error in the error range. The details of this calculation can be found in the supplement.

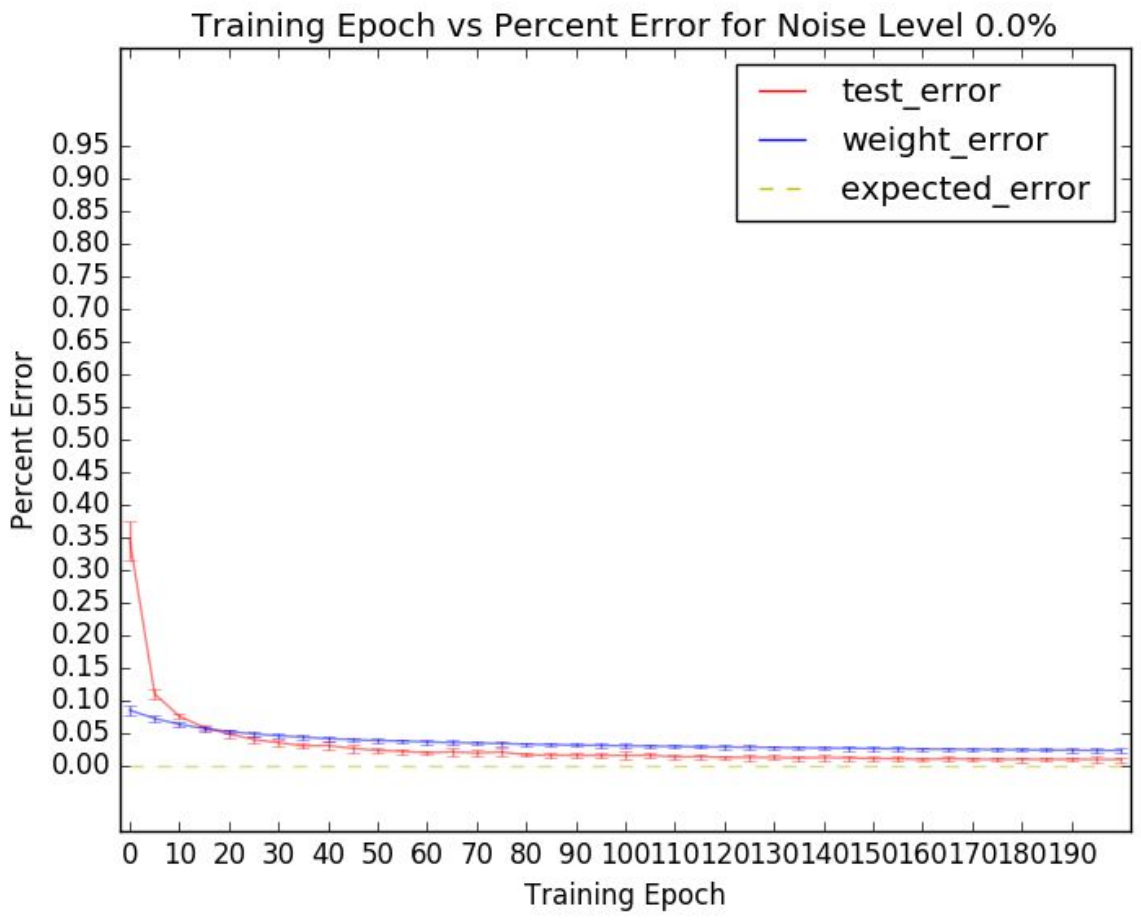

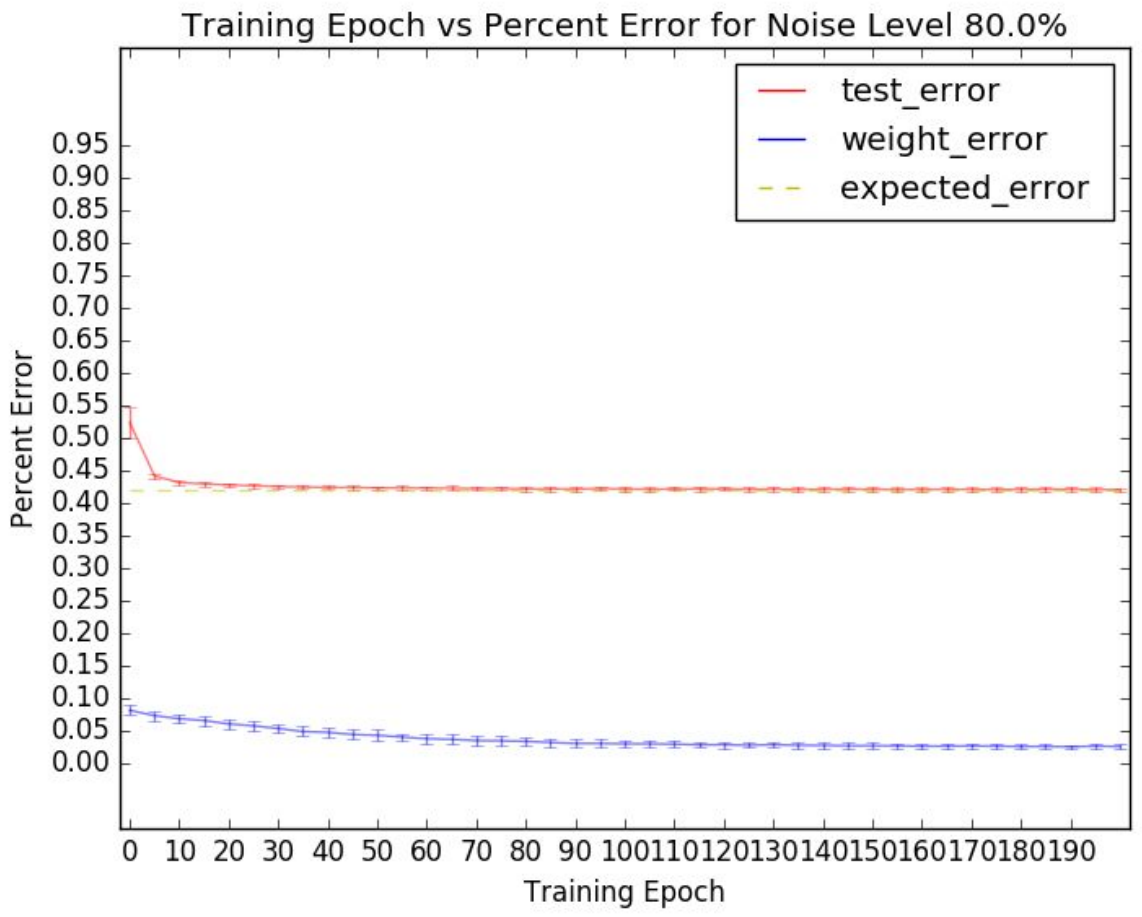

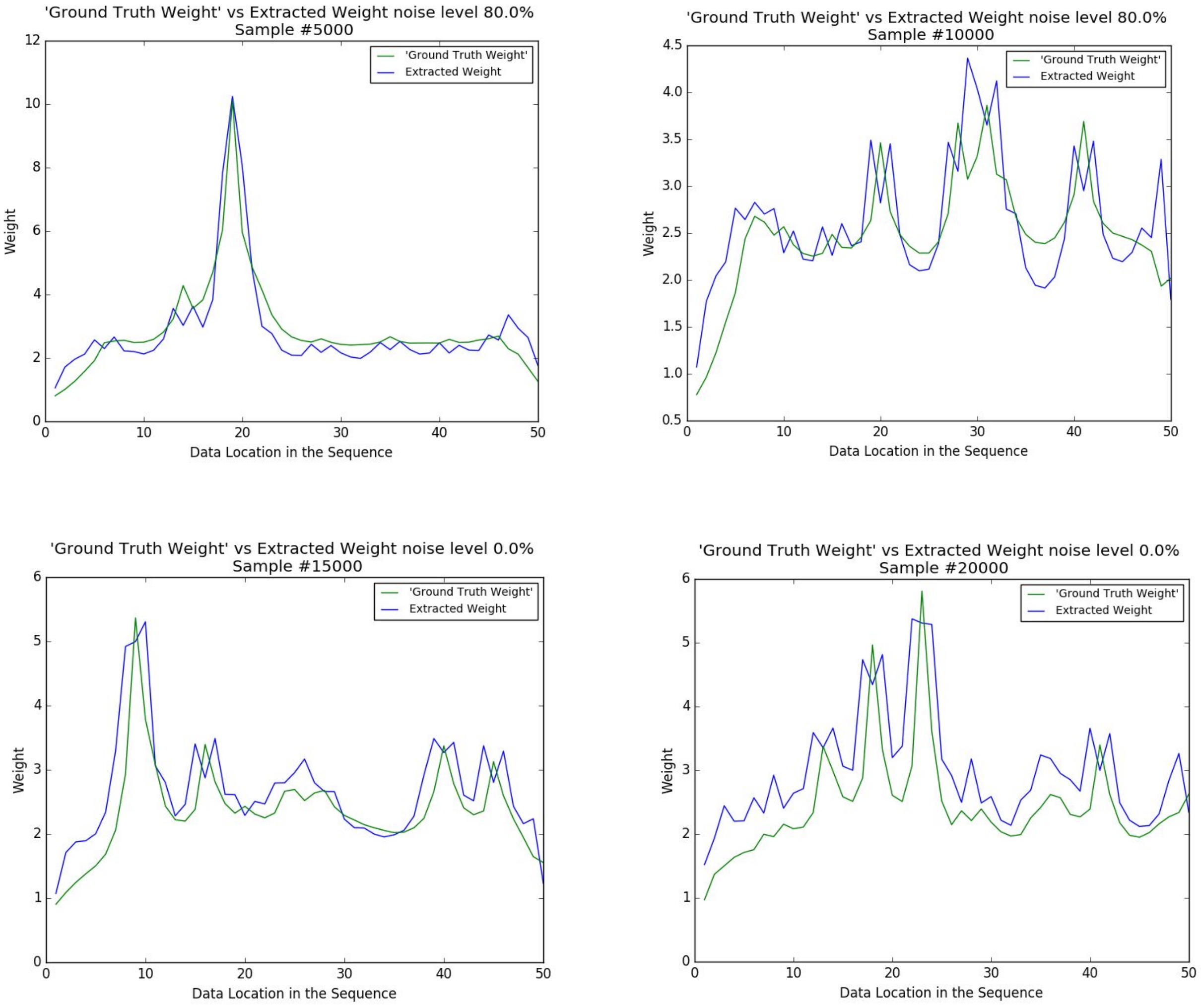

**Figure 2.** Performance assessment of contextual regression on simulated dataset- Top: Contextual regression performance on the testing set under noise level 0% and 80%. Height of the error bar is 1 standard deviation of RMSE or RMSCD among 20 runs. In each epoch (training cycle), the model was trained on 10% of the randomly selected data from the training set and tested on the whole testing set. We chose 10% of the data as one epoch, other than 100% that is ordinarily used, to show the evolution of error. Middle: Sample plot of "ground truth" context weight (green) vs context weight (blue) calculated by the embedding net in the contextual regression model for ±80% noise dataset. Bottom: Same comparison plots for ±0% noise dataset.

Under all noise levels, our model achieved high performance in both prediction accuracy and rule extraction fidelity on the testing set (Figure 2 and 1S), with RMSE less than 1% away from the expected error and RMSCD below 0.03. (Table 1S). We also visually inspected (Figure 2) some randomly selected "ground truth" context weights (green) vs context weights (blue) produced by the embedding network under noise level ±0% and ±80% (Figure 2). They are indeed visually similar which agrees with the overall low RMSCD values between the "ground truth" and network output context weights.

In certain scientific experiments, the signal to noise ratio can be extremely low. This is especially the case in some experiments of physics[20,21], social science[22,23] and biological science[24–26]. Thus we also tested our model under noise level of 200% (signal to noise ratio 1:2), 500% (signal to noise ratio 1:5) and 1000% (signal to noise ratio 1:10). We found that under all three noise levels, the prediction errors still approach the expected error very fast (Figure 2S and Table 2S). However, the RMSCD is only stably decreasing under noise level of 200% and 500%. Under 1000% noise, the RMSCD value starts to slightly increase and fluctuate when the training time gets longer. This is confirmed in our visual inspection of the sample data points (Figure 3S), where the sampled extracted weights still resembles the corresponding "ground truth" weights under 200% and 500% noise but distorted under 1000% noise.

Overall, the high prediction accuracy (relative error < 0.9% in all noise levels) and high rule extraction fidelity (RMSCD < 0.05 in all noise levels smaller than 1000%) on the simulated dataset support that our algorithm is highly reliable even under high noise level. These results also suggest that our method can be applied to datasets with missing information, as long as the effect of missing information on the prediction target is similar to a random noise of uniform distribution.

**Evaluating the Contextual Regression model on DNaseI hypersensitivity data**

To examine our method in real experimental data, we applied it to predict DNaseI hypersensitivity sites (DHSs) using histone modifications in the H1 and GM12878 cell lines. We binned all the histone modification ChIP-seq and DHS-seq data at 200bp. To consider distal relationship, we included signals from 10kbp upstream to 10kbp downstream of the prediction location, i.e. 100 bins in total. In the H1 cells, there were 27 histone marks and thus the number of features reached 100 * 27 = 2700 which is too many for modeling. Hence, we arranged the features into a 100 by 27 matrix and factorized it into the tensor products of two vectors:

$$C = F \otimes D$$

where C is the 100 * 27 combination feature matrix, F is the histone feature vector of length 27, corresponding to the number of histone marks, and D is the distal feature vector of length 100, corresponding to the number of bins. This approach greatly reduced the number of parameters in the model. In the current proof of concept stage, we further simplify and speed up the model by forcing the vector D to be the same for all the regions and applying a Lasso[27] penalty on F.

| DHS Prediction Task | Pearson Correlation | Match1 | Catch1obs | Catch1imp |
|---|---|---|---|---|
| Linear Regression | 0.586 | 44.9% | 75.3% | 79.1% |
| Lasso Regression | 0.614 | 44.3% | 76.0% | 78.6% |
| Lasso Regression (with log input) | 0.537 | 47.4% | 78.3% | 82.0% |
| Contextual Regression | 0.800 | 60.5% | 90.5% | 92.2% |
| Contextual Regression (with log input) | 0.825 | 61.1% | 90.0% | 89.8% |
| LSTM Benchmark (with log input) | 0.817 | 60.9% | 89.2% | 89.5% |

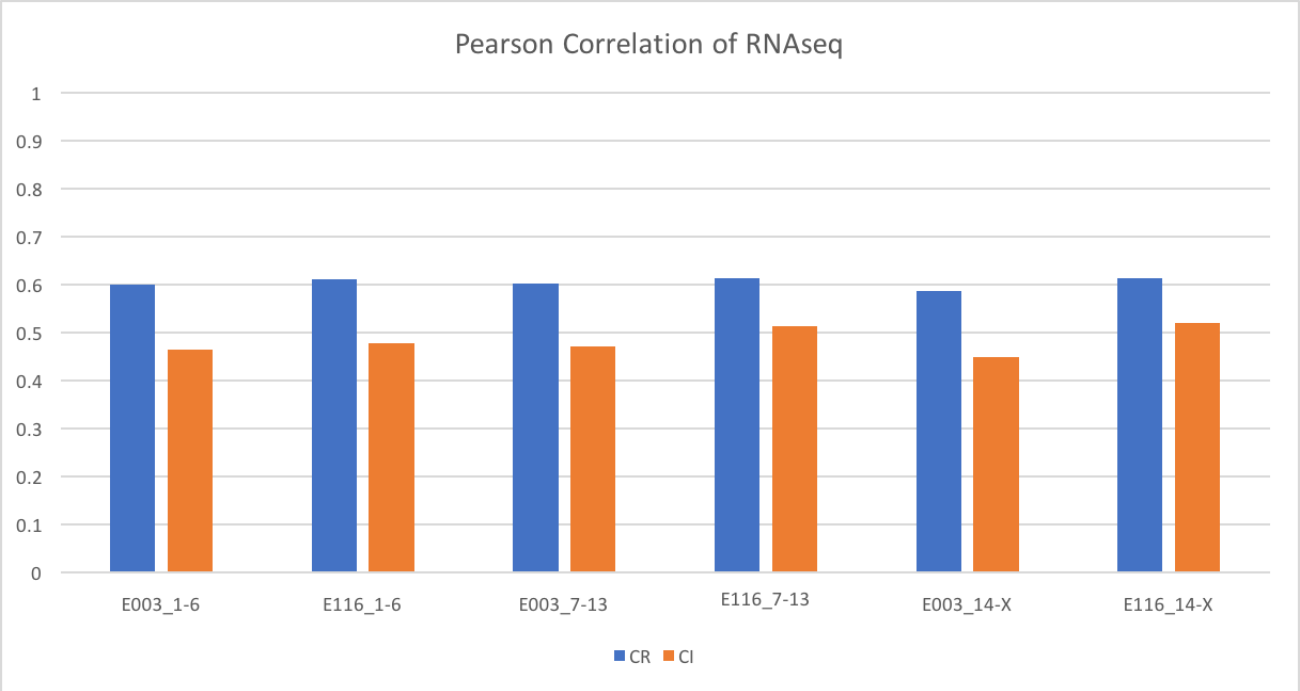

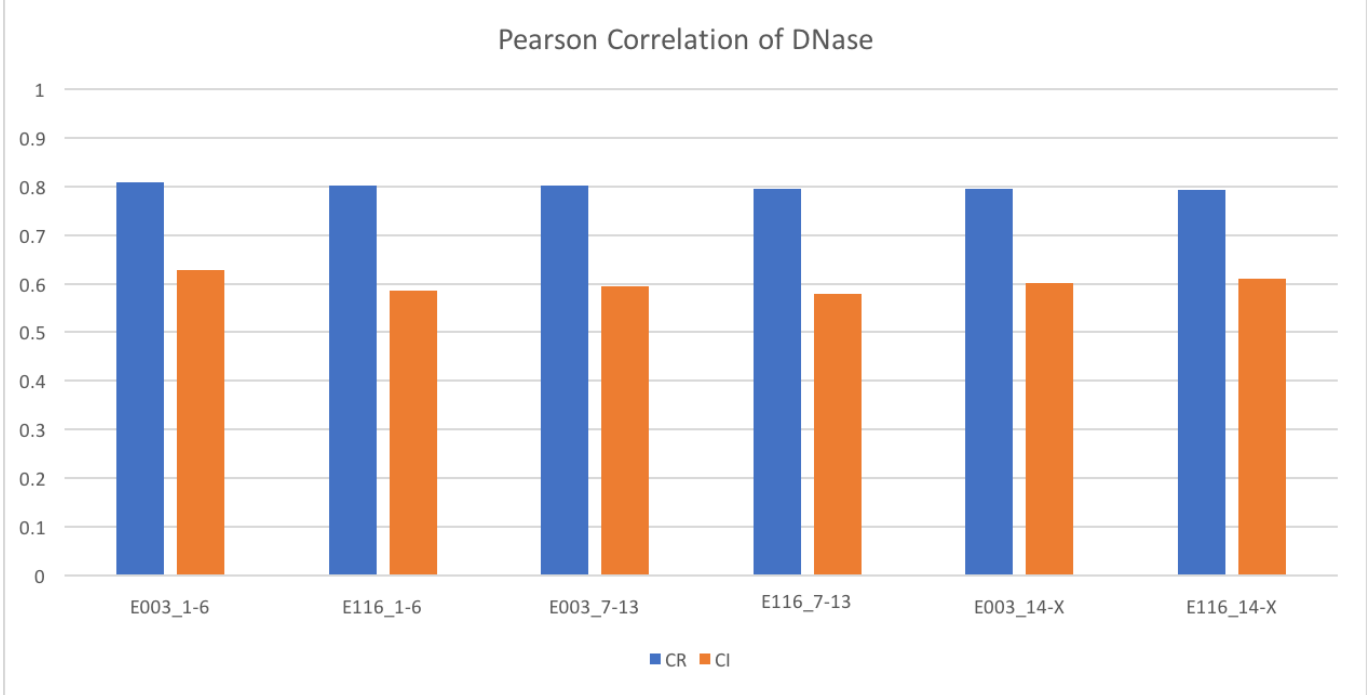

**Figure 3.** Prediction performance comparison of contextual regression with linear regression, LSTM model and ChromImpute- Top: Performance comparison of contextual regression with various linear regression and LSTM models on DHSs dataset in cell-line H1. We used four imputation quality evaluation metrics[28]: pearson correlation, match1 score (percent of 99 percentile experimental values in 99 percentile predicted values), Catch1obs (percent of 99 percentile experimental values in 95 percentile predicted values) and Catch1imp (percent of 99 percentile predicted values in 95 percentile experimental values). Bottom: Comparison of prediction accuracy (Pearson Correlation) between contextual regression (CR) and ChromImpute (CI).

We compared contextual regression with linear regressions and a benchmark neural network model implemented with Bidirectional-LSTM. The task for comparison is the prediction of DHS confidence values in cell line H1 using the 27 histone marks ChIPseq data in the same cell line (Figure 3 and 4S). The models are trained on 70% of the randomly selected genomic regions and tested on the rest 30%. The contextual regression obviously outperformed linear regression, showing its better ability to capture nonlinear relationship between histone marks and open chromatin. At the same time, its performance is as good as the LSTM benchmark, showing that our interpretation method preserves the accuracy of a complex nonlinear model in this task.

Since applying log to the histone mark data improves the result slightly, we used this technique in all the studies in this section. We then assessed the performance of contextual regression on predicting RNAseq and DHS data in cell lines H1 and GM12878 by comparing to ChromImpute[28], the state of the art imputation method based on random forest algorithm. We ran both methods in three rounds of cross validations: (1) train: chr1-6, test: 7-X, (2) train: chr7-13, test: 1-6+14-X and (3) train: 14-X, test: 1:13. The models were trained using only the data from the same cell line. Contextual regression outperformed ChromImpute on both DHS and RNAseq predictions (Figure 3 and 5S).

We also performed several other evaluations of contextual regression (See supplement for detail) on: high contribution feature validity, context weight assignment consistency and whether the target can be predicted well using only the high contribution features. As expected, the high contribution histone marks for both RNA-seq and DHS concord with previous research. Our model also assigned consistent weights to the features and made accurate prediction using only the high contribution features. These results have demonstrated that our method is not only able to achieve better performance than the state of the art models but robustly select important features as well.

**Discovering Important Histone Mark Patterns in the Open Chromatin Regions of H1:**

Next, we modified the model to search for both distal and histone mark features in the open chromatin regions. We made three changes to the model and training process: 1 we removed the restraint on distal feature vector D so that our model produces different D from different feature vectors input, 2 we did not apply log to make the result closer to the representation of the original magnitude, and 3 we applied a softmax[29] constraint on D rather than the Lasso penalty on F to force the distal weights to be concentrated on a small number of locations. To make sure the patterns we extract are valid, we checked the prediction accuracy on the training and testing set and indeed, the model can still perform prediction accurately (Table 3S).

When analyzing the histone mark contribution patterns, we extracted successfully predicted peak regions (logpval > 3 for both experimental and predicted) from both training and testing set. We applied element-wise multiplication to the context weights (100 * 27) with the magnitude of the corresponding input (also 100 * 27), normalized them into vectors (of dimension 100 * 27 = 2700) of length 1, applied pca (of dimension 2700), picked the top 100 pcs (which captured > 90% variance in all cell lines) and then clustered it with K-means clustering[30]. With k = 20, the clusters are well separated indicating by the great difference between in and across cluster distance from the cluster center (Figure 6S).

Among the cluster centers, we observed 4 types of major patterns: 1 the central dominant histone marks (with small contribution from other marks), 2 the spread histone marks, 3 the central dominant H2A.Z, and 4 the H3K27me3 pattern. Type 1 pattern (Figure 7S) has the peak contribution exactly from the center with a radius of 300bp by the H3K4me1/2/3 and H3K27ac. This type is composed of majority of the clusters (0, 1, 2, 3, 5, 6, 7, 8, 9, 12, 13, 17, 19). On the other hand, Type 2 patterns (Figure 8S) also have long range contributions (> 300bp) besides the central dominant peaks. This type is composed of cluster (11, 14, 15, 16). Type 3 pattern (Figure 9S) has central dominant (< 300 bp) contribution from H2A.Z, a modified histone. This type is composed of cluster 10 and 18. Type 4 pattern (Figure 12S), most interestingly, is highly unique compared to the previous three types. It has most significant contribution from H3K27me3 and ancillary contribution by H2A.Z, H4K8ac, H3K18ac and the H3K4me1/2/3. The effect of H3K27me3 is long range in this pattern and almost covers the whole 20kbp region. This type compose of cluster 4, a total of 3959 (4.2%) predictable DNase strong signal (logpval > 3 for both experimental and predicted) regions.

Most of the open chromatin contributors found by our method, H3K27ac[31,32], H2A.Z[33,34], H3K4me1/2/3[35], H4K20me1[36] and K4K8ac[37], have been repeatedly reported by previous research and thus are not surprising. However, H3K27me3 is not widely known to be associated with open chromatin, and their correlation with open chromatin is only mentioned in several papers[38,39]. To ensure this pattern is not caused by artifact, we visually inspected a couple of regions of type 4 pattern. All these regions have strong H3K27me3 signal that covers the whole region (Figure 10S). This is consistent with the contribution pattern (Figure 12S) discovered by our algorithm. We also found that the regions we observed contain transcription starting regions marked by the GENCODE V7 and Ref-seq annotations. After calculating the transcription start site content (with GENCODE data) around these regions, we found that about 64.4% of the regions in this cluster contain transcription start site in 3kbp radius and 78.2% in 5kbp radius (Table 4S). This is consistent with previous research[40] which reports the bivalent domains that are stem cell specific: regions contain H3K4 and H3K27 methylation sites of size 1kbp-18kbp and overlap with genes and transcription start sites. This evidence further validates the findings of our algorithm.

**Discovering Important Histone Mark Patterns in the Open Chromatin Regions of Other Cell-lines:**

We then apply our method to the DHS data in other cell-lines (H9, IMR90, GM12878, HUVEC) to search for cell line specific patterns. We found that pattern type 1-3 appear in all 5 cell-lines, however, type 4, the H3K27me3 pattern, is only found in H1 and H9, both are embryonic stem cells. This coincides with the previous finding that the correlation between H3K27me3 and open chromatin is highly cell specific[38] and mostly found in embryo cells[40,41].

We next investigated whether there are specific DNA sequence motifs associated with each pattern. We compared the appearance frequency of motifs in each cluster and in all the clusters. The p-value in the comparison was calculated using One-Proportion Z-test. We limited our discussion only to cell-lines H1 and H9, which has histone mark data for non-traditional marks.

The cluster 4 and 16 of H1 and cluster 0, 4, 5, 8, 15, 16 and 19 of H9 contain motifs that have frequency difference above 10%. H9-8 and H9-19 are classic H3K27ac patterns. Both clusters have very similar motif enrichment in top 7 (Table 5S). Their enriched motifs agree except for NKX21 and PO3F1. Among the motifs enriched in both clusters, PO5F1 and SOX2 form a complex and regulate the expression of embryonic genes[42]. PO2F1/2 activate the genes that codes for H2B proteins[43]. PO3F2 is a protein involved in differentiation[44]. We also examined the H3K27ac dominant clusters in H1 (cluster 1 and 6) and found that many of the above motifs are also significantly enriched by the standard of pval (Figure 11S and Table 6S), however less so probably due to the difference of cluster purity. Thus a consistent motif enrichment pattern is confirmed for H3K27ac dominant open chromatin regions.

H1-cluster 4 and H9-cluster 4 are both H3K27me3 patterns (Figure 12S). Their motif enrichment profile are also very similar. Specifically, in H9-cluster 4, the frequency enrichment that is highly significant are shown in Table 1.

| H9-Cluster 4 | Frequency Diff | p-value | Frequency in Cluster | Background Frequency |
|---|---|---|---|---|
| MBD2 | 26.8% | 0.0 | 89.7% | 62.9% |
| E2F1 | 25.6% | 0.0 | 83.2% | 57.6% |
| E2F4 | 24.3% | 0.0 | 63.5% | 39.2% |
| NRF1 | 23.8% | 0.0 | 75.9% | 52.0% |
| E2F3 | 21.5% | 0.0 | 37.1% | 15.6% |
| E2F2 | 15.0% | 1.75e-258 | 31.5% | 16.5% |

**Table 1.** Motif enrichment profile of cluster 4 in H9 (H3K27me3 pattern), pval of 0 values are caused by number underflow since the pval is smaller than machine accuracy

In this cluster, MBD2 is the most enriched motif (62.9% overall -> 89.7% in cluster 4) and it plays the role of a transcription repressor that can recruit deacetylase and methyltransferase[44–48]. Another class of motifs that are enriched are the E2 family factors (E2F1-4) which are very important for the cell cycles[49,50]. These functions are highly correlated with the role of stem cell: these regions present in the differentiation process and can be an example of repressor induced chromatin remodeling[51].

More interestingly, we have H9-cluster 15 and 16, which are H3K18ac dominant clusters (Figure 4). The major contribution of H3K18ac is around 200bp up or downstream. The correlation of this mark with open chromatin is previously undiscovered and our algorithm only detected it in H9 among the 5 cell lines we studied. They compose 9.7% of the total predictable peak regions. To confirm the finding, we manually inspected some of the regions in the genome browse and H3K18ac does show up strongly around peaks of its prediction, although in some cases, the other marks also co-appear at the neighboring locations (Figure 4 and 13S).

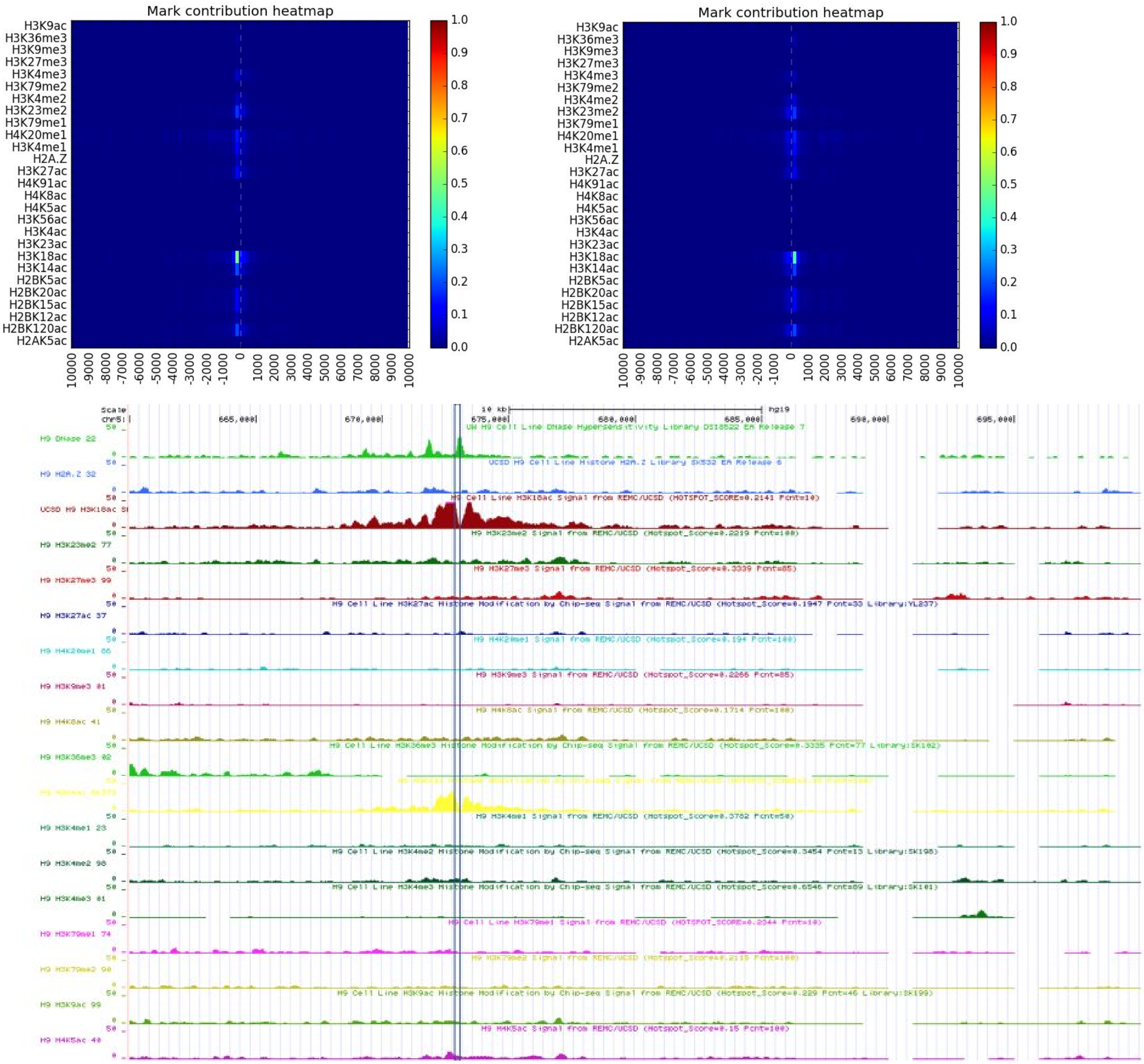

| H9-Cluster 15 | Frequency Diff | p-value | Frequency in Cluster | Background Frequency |
|---|---|---|---|---|
| USF2 | 13.1% | 2.03e-93 | 70.8% | 57.7% |
| MYC | 12.9% | 4.28e-88 | 60.3% | 47.4% |
| MAX | 12.1% | 1.71e-77 | 61.9% | 49.9% |
| MLXPL | 11.5% | 1.34e-76 | 75.5% | 64.1% |
| PLAL1 | 11.3% | 2.70e-84 | 82.4% | 71.1% |
| MYCN | 11.1% | 6.18e-70 | 48.6% | 37.5% |
| ZIC3 | 10.4% | 1.88e-65 | 77.1% | 66.7% |
| USF1 | 9.9% | 2.31e-53 | 53.2% | 43.3% |

**Figure 4.** Study of H3K18ac pattern- Top: Mark contribution patterns of cluster 15 and 16 in H9 (H3K18ac pattern). Middle: Example genome browser view of the cluster 15 (H3K18ac dominant cluster). Bottom: Motif enrichment profile of cluster 15 in H9 (H3K18ac pattern).

On the other hand, the motifs in the region of cluster 15 and 16 show a unique enrichment profile (Figure 4) among all the regions we examined. Among the enriched motifs, USF1,2 are associated with gene activation[52]. MYC family (MYC, MYCN, MAX and MLXPL) can form dimers and binds to E-box that regulates cell proliferation and differentiation[53] while PLAL1 also facilitates the same function[54]. ZIC3 works on early body axis formation[55]. All these factors fit the stem cell nature of H9 cell line.

H9-cluster 5 is H3K23me2 dominant (Figure 5), which is another previously undiscovered and H9 unique pattern. The major contribution of H3K23me2 is around 200bp downstream. It compose 5.6% of the total predictable peak regions. However, the visual correlation of H3K23me2 with open chromatin is not as strong as the dominant marks in aforementioned clusters and co-appears with other marks in the open chromatin regions (Figure 5 and 14S). This lower correlation is also found statistically by our model, which shows less percentage of contribution from H3K23me2 (max value 43.8%) than the dominant marks in other clusters (for instance, max value of H3K18ac is 112%, > 100% since there are negative contributions from some other marks).

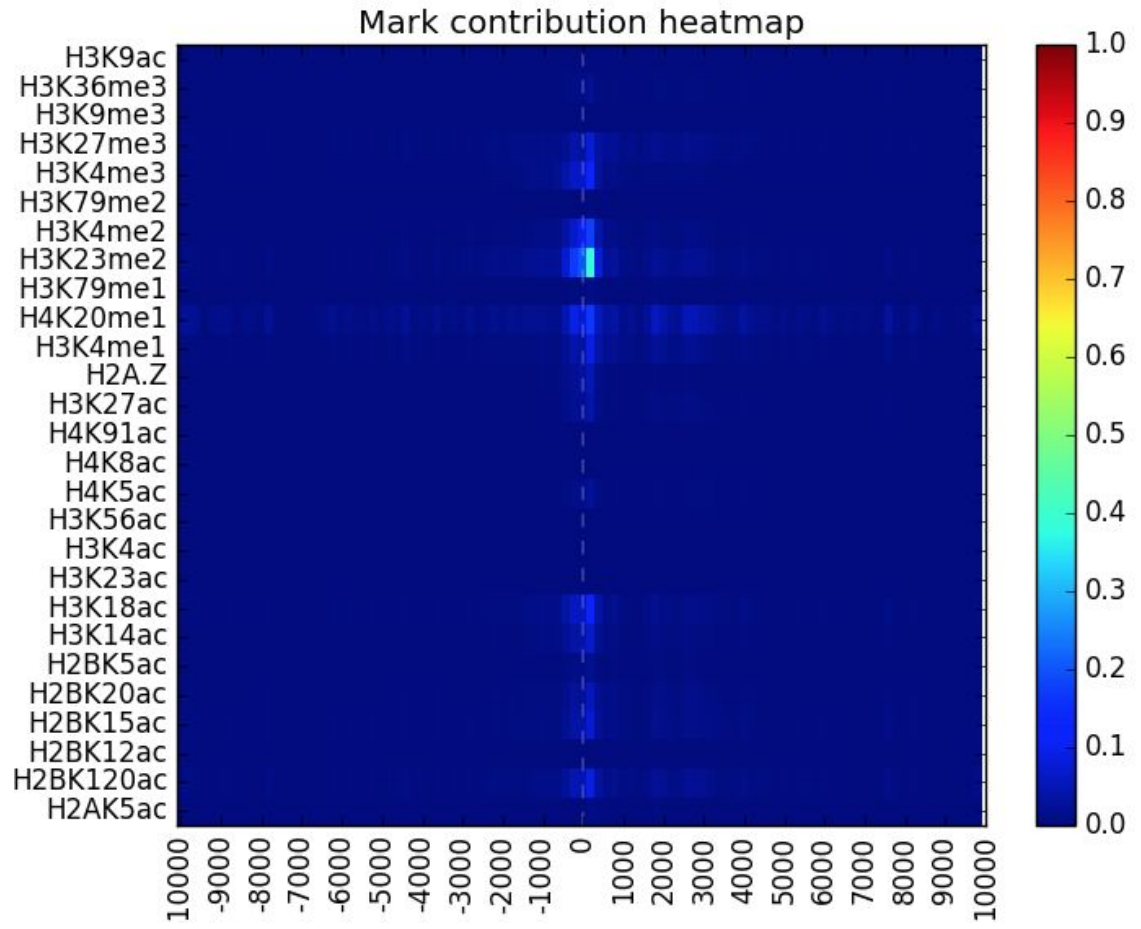

| | Cluster 5 | Cluster 4 | BG |
|---|---|---|---|
| CEBPD | - 4.1% | - 20.3% | 85.3% |
| CEBPG | - 5.3% | -16.7% | 79.7% |
| CEBPB | - 5.6% | -16.7% | 78.5% |
| E2F1 | +9.4% | +25.6% | 57.6% |
| E2F4 | +7.9% | +24.3% | 39.2% |
| E2F3 | +4.7% | +21.5% | 15.6% |
| E2F2 | +3.8% | +15.0% | 16.5% |

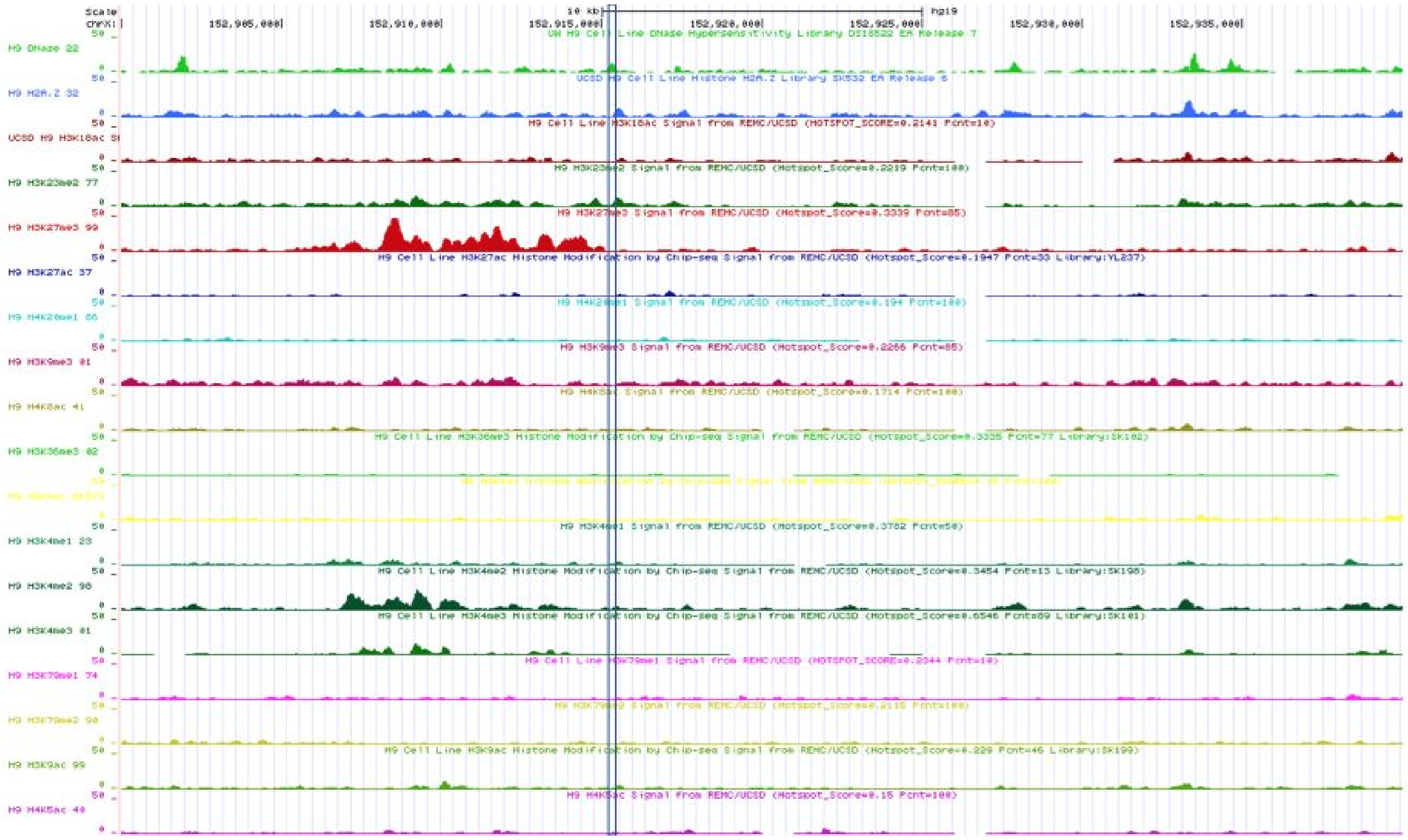

**Figure 5.** Study of H3K23me2 pattern- Top Left: Mark contribution pattern of cluster 5 H9 (H3K23me2 pattern), Top Right: Motif enrichment profile comparison between cluster 4 (H3K27me3 dominant) and 5 (H3K23me2 dominant) in H9. The 2nd and 3rd column are the motif frequency difference from the background and the 4th column is the BG (background) frequency, Bottom: Example Genome browser view of the cluster 5 (H3K23me2 pattern)

The signature motif profile of H3K23me2 is very similar to the one of H3K27me3, with high MBD2 and E2 family enrichment. The largest motif frequency difference (Figure 5) is that the H3K23me2 regions has CEBPD motif frequency close to background level compared to the H3K27me3 cluster (81.2% vs 65.0%). CEBPB and CEBPG motif frequencies also show large difference between these two clusters. Function-wise, CEBPD is associated with immune response, immune cell activation and differentiation[56]. Other motifs that have significantly different profile among these two clusters is the E2 family. One possibility is that H3K23me2 is important for the stem cell development at a certain stage which is the reason why it is only enriched in certain open chromatin regions in H9.

H1-cluster 16 and H9-cluster 0 (Figure 15S) are the spread type which include contribution from a lot of histone marks while none of them are dominant (Figure 16S). These two clusters show very similar motif enrichment patterns: strong contribution from MBD2 (+18%) and E2 (+5%-10%) families and high CEBPD frequency (more than 80% in both clusters). We hypothesize that this cluster is intermediate in its motif and histone mark patterns between the H3K27me3 and H3K23me2 clusters, which suggests that these three clusters are open chromatins in repressive states at different levels.

The above observations demonstrate that the open chromatin patterns captured by our method not only accord with visual inspection but enriched in unique motifs as well. This shows the existence of diverse histone mark and motif patterns that can relate to open chromatin. Two of the patterns, H3K18ac dominant and H3K23me2 dominant have not been found in previous literatures, which shows the sensitivity of our method in the task of important feature detection.

**Discussion:**

In this paper, we described contextual regression, a generalized nonlinear model that is human interpretable. This method performs well in terms of prediction accuracy and important feature extraction in both simulated and experimental datasets. On epigenetic datasets, our method not only outperformed previous methods in terms of accuracy, but also robustly extracted important features that are aligned with previous research.

Using our method with the assistance of K-mean clustering, we not only found open chromatin patterns that have been discovered by previous research, but also new ones that exhibit signature patterns in both significant histone marks and DNA sequence motifs. H3K18ac, H3K23me2 and H3K27me3 are unfound or rarely mentioned indicators, which highlight the sensitivity of our method and the advantage of predictive models in the search of important factors. These results prove the validity of our algorithm and also emphasize the necessity of individual studies of each cell line, which will demand interpretable machine learning methods that can reduce the manual work needed in such kind of data mining tasks.

Despite its impressive performance, one potential problem of our method is the existence of alternative models in dataset with redundancy, which indeed exists in the open chromatin dataset that we have worked on in this paper. We have developed utopia penalty technique as a straightforward solution: adding a penalty term for feature weight unbalance during training. However, more research is needed to study the behavior of contextual regression to make it more stable and effective when dealing with this kind of situations.

**Reference:**

1. Seber, G. A. F. & Lee, A. J. *Linear Regression Analysis*. (John Wiley & Sons, 2012).

2. Hosmer, D. W., Jr., Lemeshow, S. & Sturdivant, R. X. *Applied Logistic Regression*. (John Wiley &

Sons, 2013).

3. Support Vector Machines and Regularization Networks. in *Template Matching Techniques in Computer Vision* 237–262

4. LeCun, Y., Bengio, Y. & Hinton, G. Deep learning. *Nature* **521,** 436–444 (2015).

5. Rokach, L. & Maimon, O. *Data Mining with Decision Trees: Theory and Applications*. (World Scientific, 2014).

6. Alipanahi, B., Delong, A., Weirauch, M. T. & Frey, B. J. Predicting the sequence specificities of DNA- and RNA-binding proteins by deep learning. *Nat. Biotechnol.* **33,** 831–838 (2015).

7. Zhou, J. & Troyanskaya, O. G. Predicting effects of noncoding variants with deep learning-based sequence model. *Nat. Methods* **12,** 931–934 (2015).

8. Wu, X., Zhang - arXiv preprint arXiv:1611.04135, X. & 2016. Automated Inference on Criminality using Face Images. *arxiv.org* (1611).

9. Kosinski, M. & Wang, Y. Deep neural networks are more accurate than humans at detecting sexual orientation from facial images. (2017).

10. Zeiler, M. D., Fergus - European conference on computer vision, R. & 2014. Visualizing and understanding convolutional networks. *Springer* (2014).

11. Zhang, Q., Wu, Y. N., Zhu - arXiv preprint arXiv:1710.00935, S. C. & 2017. Interpretable Convolutional Neural Networks. *arxiv.org* (1710).

12. Cybenko, G. Approximation by superpositions of a sigmoidal function. *Math. Control Signals Systems* **5,** 455–455 (1992).

13. Hochreiter, S. & Schmidhuber, J. *Long Short Term Memory*. (1995).

14. Shrikumar, A., Greenside, P., Shcherbina, A. & Kundaje, A. Not Just a Black Box: Learning Important Features Through Propagating Activation Differences. *arXiv [cs.LG]* (2016).

15. Ribeiro, M., Singh, S. & Guestrin, C. ‘Why Should I Trust You?’: Explaining the Predictions of Any

Classifier. in *Proceedings of the 2016 Conference of the North American Chapter of the Association for Computational Linguistics: Demonstrations* (2016). doi:10.18653/v1/n16-3020

16. Lundberg, S. & Lee, S.-I. A unified approach to interpreting model predictions. *arXiv [cs.AI]* (2017).
17. Xu, K. *et al.* Show, Attend and Tell: Neural Image Caption Generation with Visual Attention. in *International Conference on Machine Learning* 2048–2057 (2015).
18. Mikolov, T., Sutskever, I., Chen, K., Corrado, G. S. & Dean, J. Distributed Representations of Words and Phrases and their Compositionality. in *Advances in Neural Information Processing Systems 26* (eds. Burges, C. J. C., Bottou, L., Welling, M., Ghahramani, Z. & Weinberger, K. Q.) 3111–3119 (Curran Associates, Inc., 2013).
19. Roadmap Epigenomics Consortium *et al.* Integrative analysis of 111 reference human epigenomes. *Nature* **518,** 317–330 (2015).
20. Mastoridis, T. *et al.* Radio frequency noise effects on the CERN Large Hadron Collider beam diffusion. *Physical Review Special Topics - Accelerators and Beams* **14,** (2011).
21. Zakareishvili, T. Muon Signals at a Low Signal-to-Noise Ratio Environment. (2017).
22. Bollen, K., Cacioppo, J. T., Kaplan, R. M. & Krosnick, J. A. Social, behavioral, and economic sciences perspectives on robust and reliable science: Report of the Subcommittee on Replicability in Science, Advisory *…. from the National Science …* (2015).
23. Collaboration, O. S. Estimating the reproducibility of psychological science. *Science* **349,** aac4716 (2015).
24. Reiter, M. *et al.* Quantification noise in single cell experiments. *Nucleic Acids Res.* **39,** e124 (2011).
25. Brennecke, P. *et al.* Accounting for technical noise in single-cell RNA-seq experiments. *Nat. Methods* **10,** 1093–1095 (2013).
26. Grün, D., Kester, L. & van Oudenaarden, A. Validation of noise models for single-cell transcriptomics. *Nat. Methods* **11,** 637–640 (2014).

27. Tibshirani, R. Regression shrinkage and selection via the lasso: a retrospective. *J. R. Stat. Soc. Series B Stat. Methodol.* **73,** 273–282 (2011).

28. Ernst, J. & Kellis, M. Large-scale imputation of epigenomic datasets for systematic annotation of diverse human tissues. *Nat. Biotechnol.* **33,** 364–376 (2015).

29. Bishop, C. M. *Pattern Recognition and Machine Learning*. (2013).

30. Bernstein, I. H., Garbin, C. P. & Teng, G. K. Classification Methods—Part 2. Methods of Assignment. in *Applied Multivariate Analysis* 276–314 (1988).

31. Karlić, R., Chung, H.-R., Lasserre, J., Vlahoviček, K. & Vingron, M. Histone modification levels are predictive for gene expression. *Proc. Natl. Acad. Sci. U. S. A.* **107,** 2926–2931 (2010).

32. Rintisch, C. *et al.* Natural variation of histone modification and its impact on gene expression in the rat genome. *Genome Res.* **24,** 942–953 (2014).

33. Xu, Y. *et al.* Histone H2A.Z Controls a Critical Chromatin Remodeling Step Required for DNA Double-Strand Break Repair. *Mol. Cell* **48,** 723–733 (2012).

34. Lombardi, L. *Maintenance of Open Chromatin States by Histone H3 Eviction and H2A.Z.* (2011).

35. Bannister, A. J. & Kouzarides, T. Regulation of chromatin by histone modifications. *Cell Res.* **21,** 381–395 (2011).

36. Huang, J., Marco, E., Pinello, L. & Yuan, G.-C. Predicting chromatin organization using histone marks. *Genome Biol.* **16,** 162 (2015).

37. Gomez, N. C. *et al.* Widespread Chromatin Accessibility at Repetitive Elements Links Stem Cells with Human Cancer. *Cell Rep.* **17,** 1607–1620 (2016).

38. Shu, W., Chen, H., Bo, X. & Wang, S. Genome-wide analysis of the relationships between DNaseI HS, histone modifications and gene expression reveals distinct modes of chromatin domains. *Nucleic Acids Res.* **39,** 7428–7443 (2011).

39. Zhang, W., Zhang, T., Wu, Y. & Jiang, J. Open Chromatin in Plant Genomes. *Cytogenet. Genome*

*Res.* **143,** 18–27 (2014).

40. Bernstein, B. E. *et al.* A bivalent chromatin structure marks key developmental genes in embryonic stem cells. *Cell* **125,** 315–326 (2006).

41. Chen, T. & Dent, S. Y. R. Chromatin modifiers and remodellers: regulators of cellular differentiation. *Nat. Rev. Genet.* **15,** 93–106 (2013).

42. Takahashi, K. *et al.* Induction of pluripotent stem cells from adult human fibroblasts by defined factors. *Cell* **131,** 861–872 (2007).

43. Segil, N., Roberts, S. B. & Heintz, N. Mitotic phosphorylation of the Oct-1 homeodomain and regulation of Oct-1 DNA binding activity. *Science* **254,** 1814–1816 (1991).

44. Wong, W. T. *et al.* Discovery of novel determinants of endothelial lineage using chimeric heterokaryons. *Elife* **6,** (2017).

45. Hendrich, B. & Bird, A. Identification and Characterization of a Family of Mammalian Methyl-CpG Binding Proteins. *Mol. Cell. Biol.* **18,** 6538–6547 (1998).

46. Ng, H. H. *et al.* MBD2 is a transcriptional repressor belonging to the MeCP1 histone deacetylase complex. *Nat. Genet.* **23,** 58–61 (1999).

47. Fujita, H. *et al.* Antithetic effects of MBD2a on gene regulation. *Mol. Cell. Biol.* **23,** 2645–2657 (2003).

48. Cramer, J. M. *et al.* Probing the dynamic distribution of bound states for methylcytosine-binding domains on DNA. *J. Biol. Chem.* **289,** 1294–1302 (2014).

49. Gaubatz, S. *et al.* E2F4 and E2F5 Play an Essential Role in Pocket Protein–Mediated G1 Control. *Mol. Cell* **6,** 729–735 (2000).

50. Tsai, S.-Y. *et al.* Mouse development with a single E2F activator. *Nature* **454,** 1137–1141 (2008).

51. de la Serna, I. L., Ohkawa, Y. & Imbalzano, A. N. Chromatin remodelling in mammalian differentiation: lessons from ATP-dependent remodellers. *Nat. Rev. Genet.* **7,** 461–473 (2006).

52. Viollet, B. *et al.* Immunochemical characterization and transacting properties of upstream stimulatory factor isoforms. *J. Biol. Chem.* **271,** 1405–1415 (1996).

53. Meyer, N. & Penn, L. Z. Reflecting on 25 years with MYC. *Nat. Rev. Cancer* **8,** 976–990 (2008).

54. Chou, W.-Y. *et al.* Human Spot 14 protein is a p53-dependent transcriptional coactivator via the recruitment of thyroid receptor and Zac1. *Int. J. Biochem. Cell Biol.* **40,** 1826–1834 (2008).

55. Zhu, L., Zhou, G., Poole, S. & Belmont, J. W. Characterization of the interactions of human ZIC3 mutants with GLI3. *Hum. Mutat.* **29,** 99–105 (2008).

56. Tsukada, J., Yoshida, Y., Kominato, Y. & Auron, P. E. The CCAAT/enhancer (C/EBP) family of basic-leucine zipper (bZIP) transcription factors is a multifaceted highly-regulated system for gene regulation. *Cytokine* **54,** 6–19 (2011).

**Supplementary Materials:**

**Methods:**

**Setup of the Neural Networks used in This Article:**

The network structure we used for the three tasks are shown as followed:

**Contextual Regression network structure used in section: Evaluating the Contextual Regression model on simulated data**

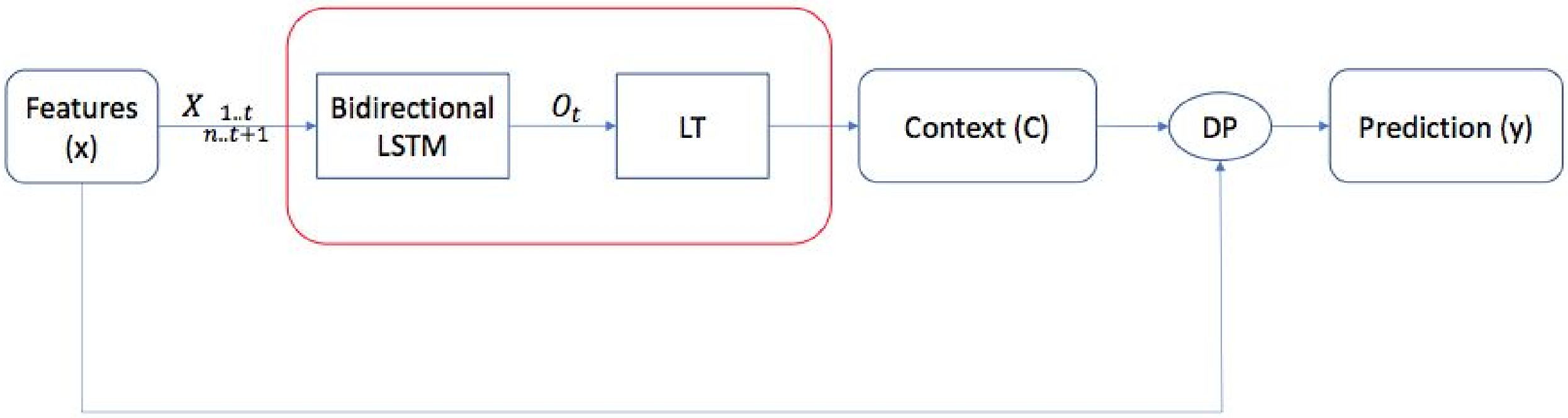

**Contextual Regression network structure used in section: Evaluating the Contextual Regression model on DNaseI hypersensitivity data**

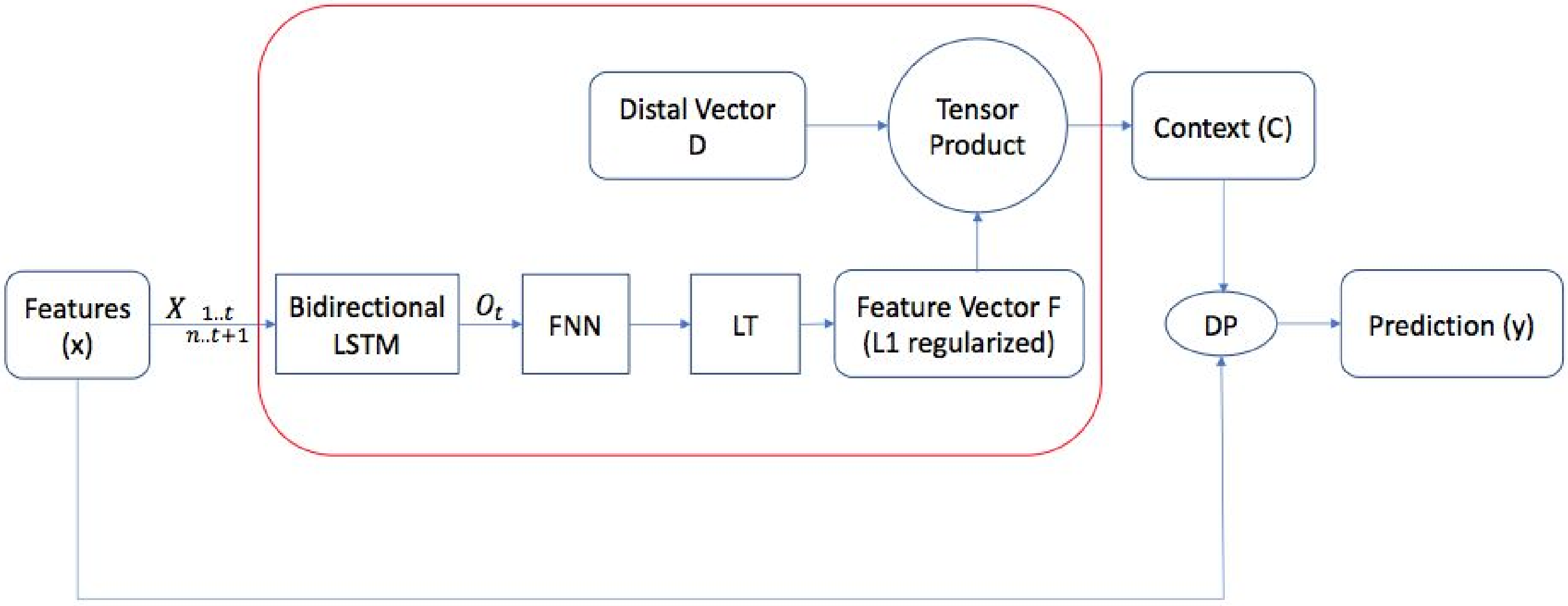

**Neural Network structure used in section as performance benchmark: Evaluating the Contextual Regression model on DNaseI hypersensitivity data**

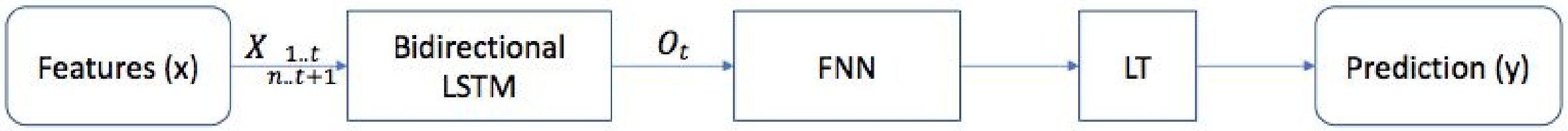

**Contextual Regression network structure used in section: Discovering Important Histone Mark Patterns in the Open Chromatin Regions**

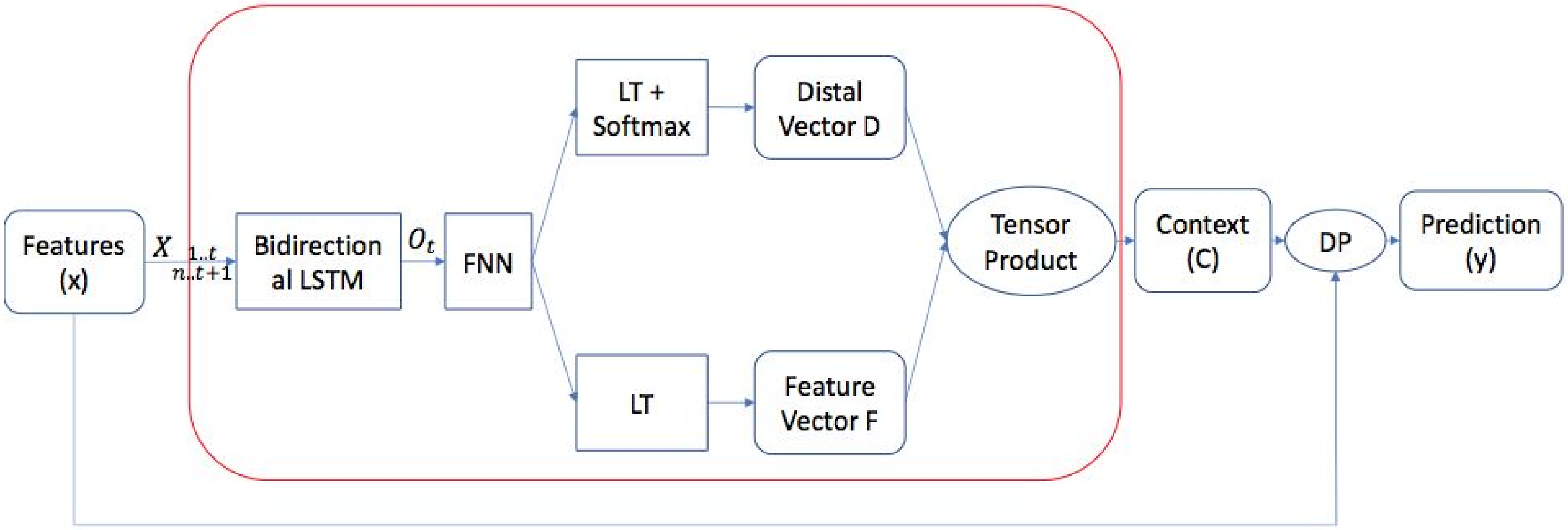

**Figure 1ST.** Architecture of the contextual regression models used in the main article, the embedding network in each model is marked by red squares.

In the graphs, features is the input vector and y is the prediction target, context is the linear model produced by the embedding network from the input vector, LT represents linear transformation which maps a vector x to Ax+b where A is an matrix and b is an vector, FNN represents Feedforward neural network which maps a vector x to s(Cx+d) where C is an matrix, d is an vector and s is the activation function (we used tanh), DP represents the dot product of the two input vectors. $O_t$ is the output of the Bidirectional LSTM at the distal location t, after processing $X_{1..t}$ (forward) and $X_{n..t+1}$ (backward) in sequence. In LT and FNN, matrices A, C and vectors b, d are automatically learned by the model to maximize the prediction accuracy.

The Bidirectional LSTM is implemented with the tensorflow static_bidirectional_rnn function using layer size of 2. The batch size is set to 10. RMSD between the real and predicted data is used as the cost function in training, alone with lasso or utopia penalty as mentioned in the main text. The ADAM optimization algorithm[1] is used to train the model, learning rate is set to 1e-4, maximum gradient norm to 5. Hidden size is set to 80 in the ground truth test, 30 in the prediction accuracy comparison task and 40 in the discovering task, the hidden sizes are chosen to be approximated to the number of features in each task.

In the task of DHS and RNAseq prediction, there are only a small number of high confidence DHS and RNAseq regions. Thus from the training set, we randomly removed 70% of the data between 60 and 90 percentile, 80% if the data between 30 and 60 percentile and 90% of the data below 30 percentile to balance the dataset.

**Setup of the Ground Truth Test (in section: Evaluating the Contextual Regression model on simulated data):**

The input of the Neural Network is a sequence of features of length 50 and the output is a float value y. Each feature ($x_1 \ldots x_{50}$) are sampled from the set (1, 3, 9, 27, 81) with probability (70%, 20%, 7%, 2.5%, 0.5%) and a random fraction below 20% is added or subtracted from the value to increase data diversity. The "Ground Truth weight" $c_i$ and the expected output y are calculated using the function in the main text. After y is generated, a random noise between ±n% (n ranges from 0 to 1000 as mentioned in the main text) is added to y to simulated a noisy data collection environment.

In each run a total of 100000 data points are created, 70% are used for training and 30% are used for testing. A total of 20 simulations are run. The difference between learned weight and expected weight are measured by root mean square cosine distance (RMSCD) with the formula below:

$$\text{RMSCD} = \sqrt{\frac{1}{N}\sum_{i=1}^{N} CD(w_i, u_i)}$$

$$CD(w_i, u_i) = 1 - \left|\frac{w_i \cdot u_i}{\|w_i\|_2 \|u_i\|_2}\right|$$

**Calculation of the Expected Error in the Simulated Data (in section: Evaluating the Contextual Regression model on simulated data):**

Under noise level of a, the error is uniformly sampled from the interval (-a, a). At a certain data point, suppose the error rate sampled is x% ∈ (-a, a). Then the target value after the addition of noise will be (1+x%) of the target value before the addition of noise, which is an x% different. Thus the expected error in RMSD will be the following integral:

$$\sqrt{\frac{\int_{-a}^{a} x\%^2 dx}{\int_{-a}^{a} (1+x\%)^2 dx}} = \sqrt{\frac{2a^3}{2a^3+6a}} = \sqrt{\frac{1}{1+\frac{3}{a^2}}}$$

**Source of Data:**

Data of DHS, RNAseq and ChIP-seq are downloaded from the roadmap website (http://www.roadmapepigenomics.org/). The ENCODE blacklist[2] is used to filter the data. The motif scanning is made using the FIMO function of MEME with HOCOMOCOv9.meme database. The transcription site and coding sequence statistics is done with GENCODE Gene V27lift37 file downloaded from the Table Browser (http://genome.ucsc.edu/cgi-bin/hgTables). The motif appearance frequency (MAF) is defined as shown below:

$$\text{MAF} = \frac{Number\ of\ 40kbp\ regions\ that\ contain\ the\ motif}{Total\ number\ of\ regions}$$

**Comparison of Contextual Regression with Linear Regression, Bidirectional-LSTM and ChromImpute (in section: Evaluating the Contextual Regression model on DNaseI hypersensitivity data):**

The linear regression is trained with the same setting and input features as contextual regression using tensorflow. The only difference is the replacement of contextual regression model with a linear regression model.

The architecture of the bidirectional-LSTM network is as illustrated in the previous section of the supplement material (Setup of the Neural Networks used in This Article). Its parameter setting is the same as the setting of contextual regression.

The ChromImpute model is run using the same setup as shown in the ChromImpute website (http://www.biolchem.ucla.edu/labs/ernst/ChromImpute/). The prediction is first done with 25bp resolution as what ChromImpute is designed for, and then the blacklisted regions are removed and the neighboring bins are averaged to yield data of 200bp resolution. The data used for ChromImpute benchmarking are downloaded directly from the processed ChromImpute data folder (https://personal.broadinstitute.org/jernst/roadmapconverted/CONVERTEDDATA).

**K-mean Clustering (in section: Discovering Important Histone Mark Patterns in the Open Chromatin Regions of H1):**

K-mean clustering is used to classify the pattern of histone marks around open chromatin regions. This task is carried out using the KMeans function in the python sklearn package with K = 20 and fixed random state = 0. Before clustering, we applied PCA on the weighted contribution of histone marks at each data point and the top 100 principal components are kept for the clustering step.

**In and Across Cluster Distance Calculation (in section: Discovering Important Histone Mark Patterns in the Open Chromatin Regions of H1):**

All the vectors are first normalized to a length of 1. In cluster distance is the average squared

RMSD between each member in the cluster and the cluster center (mean of the member vectors). Across cluster distance of cluster i and j is the average squared RMSD between each member in the cluster i and the cluster center of j.

**Feature Contribution (in section: Supplementary Information - Model Performance Check):**

Feature contribution $C_N^W$ is calculated using the formula below, this formula combines the context weight assigned by the embedding function and the feature magnitude:

$$C_N^W = \frac{C^W}{||C^W||_2}$$

$$C^W = C \odot H$$

$$H_i = h \cdot D$$

In the formula, C is the context weight of each histone mark, h is the histone mark signal strength vector with each element the signal strength at the corresponding location, D is the distal vector and $\odot$ represents element-wise multiplication.

**Supplementary Information:**

**Model Performance Check (in section: Evaluating the Contextual Regression model on DNaseI hypersensitivity data):**

A good model for data interpretation needs to be accurate but also stable in terms of the result it produces. We assessed the feature selection robustness of contextual regression during 5 runs on the H1 DHS and RNA-seq dataset. In each run, the whole dataset was divided randomly into training and testing sets of 7:3 ratio. During the similarity measurement, we calculated the average feature contribution of each histone mark in 3 data classes (detail of the calculation can be found in the previous section: Feature Contribution) : (1) the whole dataset, (2) the peak regions, defined by p-value > 3, and (3) the flat regions, defined by p-value < 1. We used cosine similarity, the cosine value of the angle between a pair of vectors, as the measure of similarity between weight vectors.

The similarity of distal weight is above 0.9 for both DHS-seq and RNAseq. The similarity of feature weights on the whole dataset and in the flat regions are relatively low, which is not surprising since most of the regions have low value signals and thus there are diverse sets of weight assignment that can predict them to be flat. In the regions of our interest, the peak regions, we observed highly consistent feature weights. For DHS, the average cosine similarity between feature weights is 0.991 and for RNAseq, the average cosine similarity between feature weights is 0.960 (Table 1ST).

| Weight_similarity | DHS | RNAseq |
|---|---|---|
| Feature weight (whole dataset) | 0.262 | 0.017 |
| Feature weight (flat regions) | 0.310 | 0.025 |
| Feature weight (peak regions) | 0.991 | 0.960 |
| Distal weight | 0.963 | 0.918 |

**Table 1ST.** The average cosine similarity of feature contribution vectors among 5 runs

| Avg_weight | DNase |
|---|---|
| H2A.Z | 0.874 |
| H4K8ac | 0.186 |
| H4K20me1 | 0.162 |
| H3K27me3 | 0.075 |
| H3K36me3 | 0.058 |
| H3K27ac | 0.043 |
| H3K9me3 | 0.030 |
| H3K4me1 | 0.026 |
| H3K18ac | 0.024 |
| H4K91ac | 0.024 |
| H3K9ac | 0.022 |
| H3K79me1 | 0.020 |
| H3K4ac | 0.019 |
| H3K79me2 | 0.019 |
| H2AK5ac | 0.019 |
| H2BK12ac | 0.018 |
| H3K23ac | 0.017 |
| H3K14ac | 0.014 |
| H4K5ac | 0.014 |
| H2BK15ac | 0.013 |
| H2BK5ac | 0.013 |
| H2BK20ac | 0.011 |
| H2BK120ac | 0.010 |
| H3K4me2 | 0.008 |
| H3K56ac | 0.007 |
| H3K4me3 | 0.005 |
| H3K23me2 | 0.003 |

| Avg_weight | RNAseq |
|---|---|
| H3K36me3 | 0.732 |
| H4K8ac | 0.387 |
| H2A.Z | 0.218 |
| H4K20me1 | 0.176 |
| H3K4me3 | 0.082 |
| H3K4me2 | 0.073 |
| H3K9ac | 0.059 |
| H3K27ac | 0.055 |
| H3K4me1 | 0.049 |
| H3K18ac | 0.033 |
| H2BK15ac | 0.027 |
| H4K91ac | 0.025 |
| H2BK12ac | 0.023 |
| H3K23ac | 0.023 |
| H2AK5ac | 0.021 |
| H4K5ac | 0.021 |
| H2BK20ac | 0.021 |
| H2BK5ac | 0.021 |
| H3K79me2 | 0.018 |
| H3K23me2 | 0.017 |
| H3K79me1 | 0.017 |
| H3K9me3 | 0.013 |
| H3K56ac | 0.013 |
| H2BK120ac | 0.012 |
| H3K14ac | 0.008 |
| H3K4ac | 0.007 |
| H3K27me3 | 0.004 |

**Table 2ST.** Feature mark contribution for DNase and RNAseq prediction task, ranked from high to low

Among the top 10 features (Table 2ST) with the largest feature contribution in DHS prediction, 5 of them (H3K27me3, H3K27ac, H3K9me3, H3K4me1 H3K36me3) are core marks[3]. Besides, H2A.Z[4,5], H4K20me1[6] and H4K8ac[7] have been found to be highly associated with open chromatin formation. In RNAseq, on the other hand, the transcription mark H3K36me3[8,9] is ranked number 1 as expected. H2A.Z, H4K20me1, H3K4me1/2/3 and H3K27ac[10,11] are also selected in the top 10, which confirmed that marks highly ranked by our model match the result of analysis in the aforementioned papers. In open chromatin, however, H3K4me2/3 are not important contributors. We found that these two marks have relatively low average signal strength in the DNase peak regions (contain location with logpval > 3) after we applied log to the marks (Table 3ST). This may explain why they do not have a high importance ranking in our model: the Lasso penalty will prefer features with higher magnitude to reduce the weight penalty and thus give less emphasis to the features that have small magnitudes.

| Mark in Region | Avg signal |
|---|---|
| H2A.Z | 0.436 |
| H4K8ac | 0.431 |
| H4K20me1 | 0.371 |
| H3K27ac | 0.363 |
| H3K18ac | 0.351 |
| H3K36me3 | 0.342 |
| H3K4me1 | 0.327 |
| H3K9me3 | 0.321 |
| H3K9ac | 0.315 |
| H3K4me2 | 0.31 |
| H3K27me3 | 0.31 |
| H4K91ac | 0.292 |
| H3K79me2 | 0.292 |
| H2BK20ac | 0.287 |
| H2BK15ac | 0.283 |
| H3K79me1 | 0.282 |
| H2BK5ac | 0.28 |
| H2BK12ac | 0.278 |
| H3K4ac | 0.276 |
| H2AK5ac | 0.273 |
| H3K14ac | 0.273 |
| H3K23ac | 0.269 |
| H4K5ac | 0.267 |
| H2BK120ac | 0.254 |
| H3K23me2 | 0.251 |
| H3K56ac | 0.238 |
| H3K4me3 | 0.237 |

**Table 3ST.** Average signal strength of histone marks in the DNase peak region (contain location with logpval > 3)

To check the predictivity of the marks selected by our model, we predict DNase and RNAseq using only the top 10 and top 18 features and found that the prediction accuracy is largely preserved (Table 4ST).

| | Correlation | Match1 | Catch1obs | Catch1imp |
|---|---|---|---|---|
| Dnase_all | 0.817 | 60.6% | 89.9% | 89.7% |
| Dnase_top_18 | 0.783 | 56.6% | 86.2% | 88.1% |
| Dnase_top_10 | 0.745 | 53.0% | 81.8% | 85.8% |
| RNAseq_all | 0.622 | 38.4% | 77.2% | 76.8% |
| RNAseq_top_18 | 0.649 | 35.1% | 75.1% | 75.4% |
| RNAseq_top_10 | 0.587 | 34.1% | 72.8% | 73.1% |

**Table 4ST.** Prediction accuracy of contextual regression using only selected histone marks

**Possibility of Alternative Model (in section: The contextual regression method):**

Consider the case where regression function g is the identity function. Then the formula for the loss function of y and $c_i$ are:

$$L_y = \|y - \hat{y}\|_2 = \left\| \sum_i (c_i - \hat{c}_i)\, x_i \right\|_2$$

$$L_{c_i} = \|c_i - \widehat{c_i}\|_2$$

Where y and $c_i$ are the prediction target and the i[th] element of the context weight vector, and $\widehat{y}$ *and* $\widehat{c_i}$ are their predicted counterparts generated by the contextual regression model. Their derivative to a model parameter $\theta_j$ are:

$$\frac{\partial L_y}{\partial \theta_j} = \frac{\partial L_y}{\partial \widehat{y}} \frac{\partial \widehat{y}}{\partial \theta_j} = \frac{\partial L_y}{\partial \widehat{y}} \sum_i \frac{\partial \widehat{c_i}}{\partial \theta_j} x_i$$

$$\frac{\partial L_{c_i}}{\partial \theta_j} = \frac{\partial L_{c_i}}{\partial \widehat{y}} \frac{\partial \widehat{y}}{\partial \theta_j} = \frac{\partial L_{c_i}}{\partial \widehat{y}} \sum_i \frac{\partial \widehat{c_i}}{\partial \theta_j} x_i$$

Thus $\frac{\partial L_{c_i}}{\partial \theta_j}$ will have the same sign as $\frac{\partial L_y}{\partial \theta_j}$ only if $\frac{\partial L_y}{\partial \widehat{y}}$ and $\frac{\partial L_{c_i}}{\partial \widehat{y}}$ has the same sign. This is not true in general unless we optimize each $\widehat{c_i}$ separately, in which case $L_y = L_{c_i} |x_i|$. This shows that when we move $L_y$ in its descent direction, each individual $L_{c_i}$ may not be moved in its descent direction as well. Therefore, when $L_y$ is at its local minimum, it is not necessarily true that $L_{c_i}$ is at its local minimum as well.

**Utopia Penalty Technique (in section: The contextual regression method):**

We have shown that contextual regression is highly effective in terms of extracting important features. However, due to the high flexibility of neural network, it may only find one of the few alternative models that can give similar prediction results. Sometimes the features are not completely informationally independent, especially when people are putting in every features that possibly have an effect. This problem is even more cumbersome to solve for other neural network interpretation methods which have interpretation processes that are decoupled from the training processes.

To mitigate this problem, methods such as PCA[12], gaussian elimination[13] can be pre-apply to simplify the data. Reducing the number of neurons in the network to shrink the possible model space is also a good option. Besides these traditional approaches, we would like to introduce a more straight-forward way through the mean of modifying the cost function which allows users to monitor the model lively during training.

In our previous section, we applied the Lasso penalty to make the weight sparse. This approach will "deprive" weight from the "less effective" features, which also have predictive power but can be replaced by the combination of other features that require smaller weight assignment. We can counteract this effect by an opposite term that penalizes unbalance feature weights.

One possible term of such is the cosine distance between the context weight vector and a "utopia" vector $u$ which has value 1 for all of its entries. This leads to a cost function as followed:

$$cost = \|y - \widehat{y}\|_2 + L + B$$

$$L = \lambda \|w\|_1$$

$$B = \mu \|CD(w, u)\|_1$$

$$CD(w, u) = 1 - \left| \frac{w \cdot u}{\|w\|_2 \|u\|_2} \right|$$

$$u = [1,\ 1,\ ..,\ 1]$$

Where $w$ is the context weight, $\lambda$ is the Lasso penalty parameter, and CD stands for cosine distance. In practice, the user can set their minimum accuracy tolerance and gradually tune up the utopia penalty parameter $\mu$ until the accuracy is below the tolerance threshold.

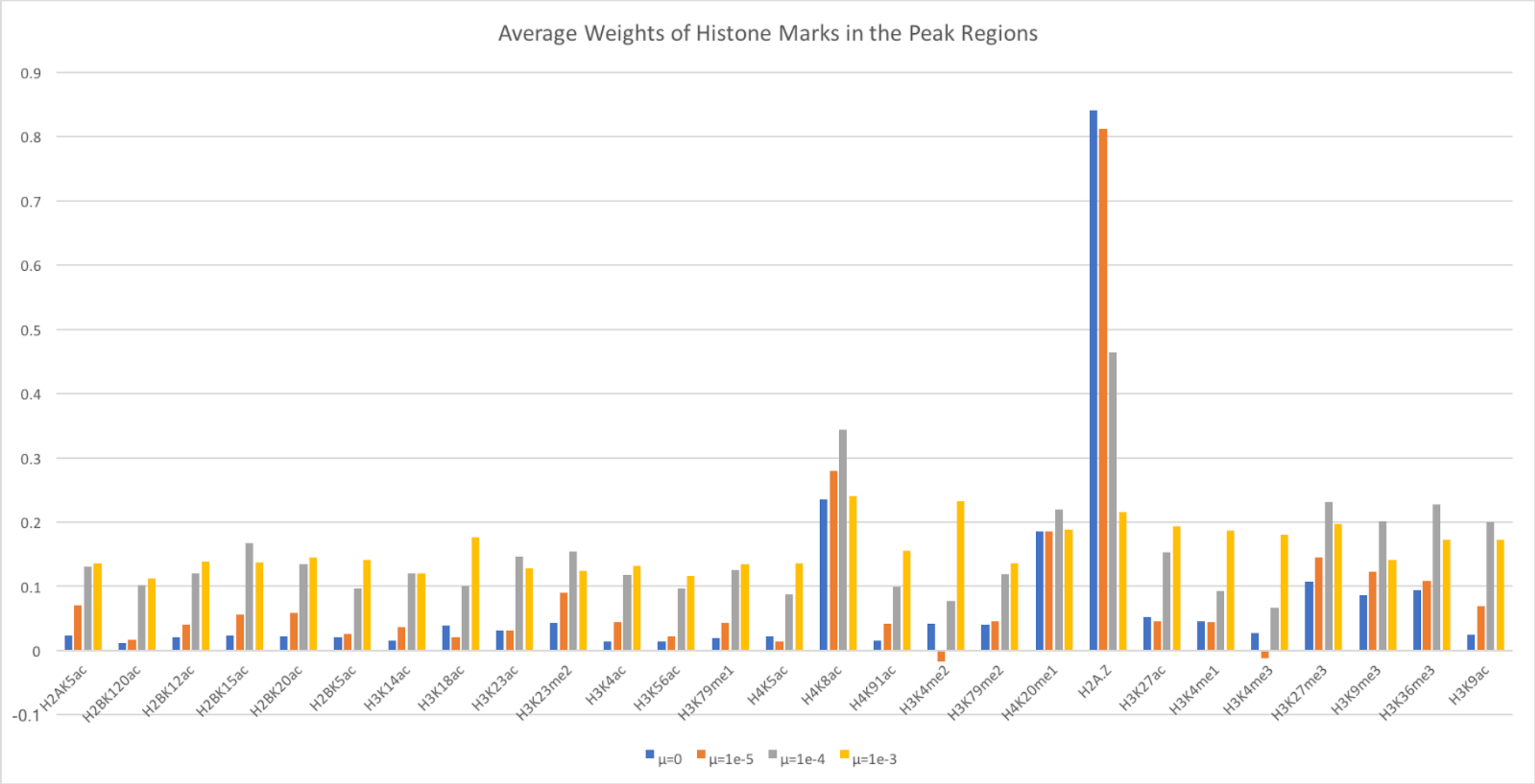

**Figure 2ST.** Average normalized contribution of histone marks to open chromatin in the peak regions under 4 difference value of utopia penalty parameter ($\mu$)

As a demonstration, we performed this strategy on H1 DNase data. We started with $\lambda$ = 1e-3 and $\mu$ = 0, increment $\mu$ from 1e-5 to 1e-1 in an exponential step size. Each run lasted for 150 epochs. The accuracy of the model with $\mu$ = 0 is 60.4% Match1 at epoch 150, thus we set the accuracy tolerance to be 59.4% Match1. We found that a utopia penalty parameter between 1e-5 and 1e-3 gives us Match1 above the tolerance. Under these levels of utopia penalty, we can see the contribution of the highest mark, H2A.Z decreases and the contribution of other marks increases in tune with the increment of utopia penalty strength (Figure 2ST). This illustrates that tuning up the utopia penalty is an effective and straightforward way to obtain diverse set of alternative models.

**Supplementary Figures:**

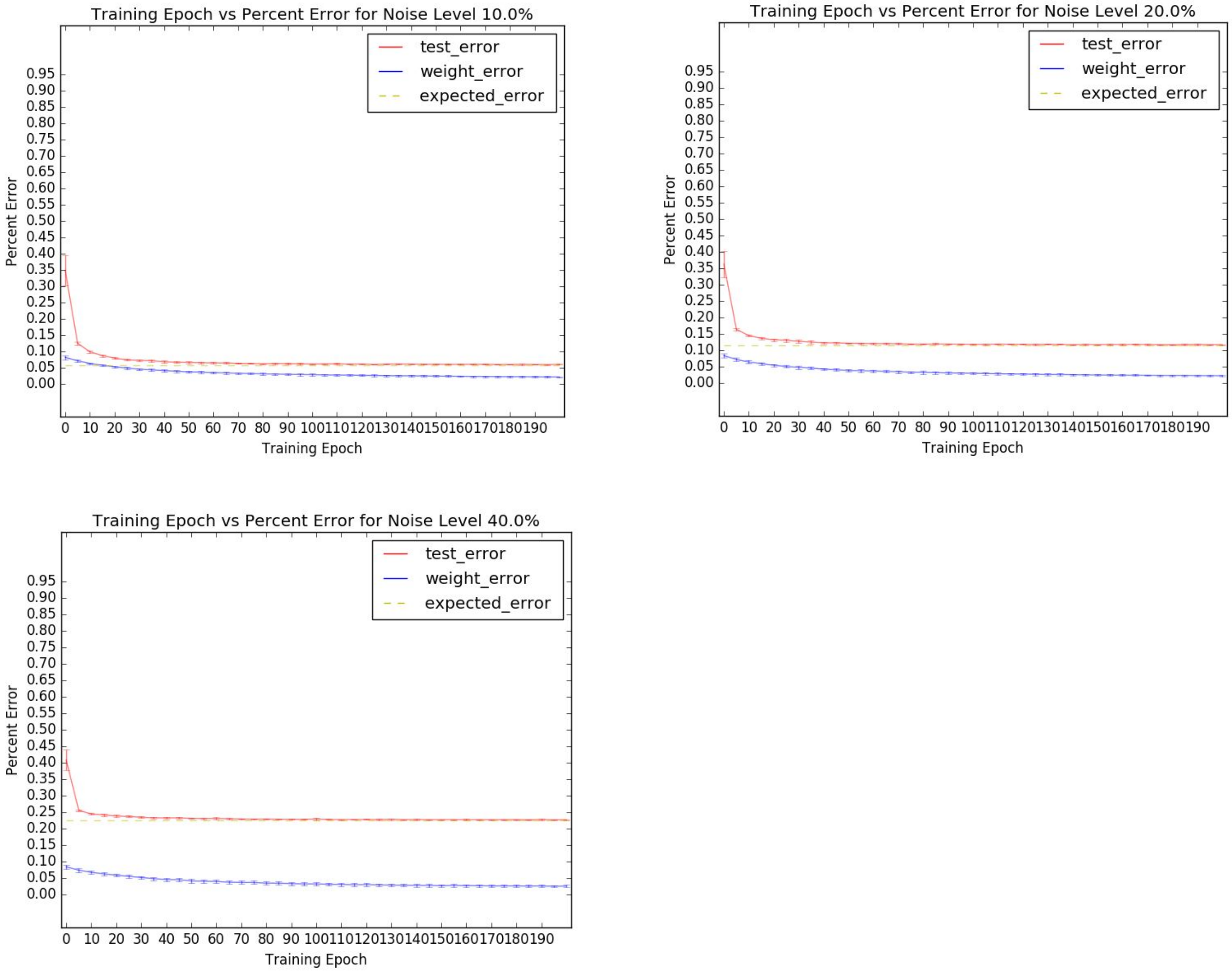

**Figure 1S.** Contextual regression performance on the simulated datasets under noise level 10%, 20% and 40%. Height of the error bar is 1 standard deviation of RMSE or RMSCD among 20 runs. In each epoch (training cycle), the model is trained on 10% of the randomly selected data from the training set and tested on the whole testing set.

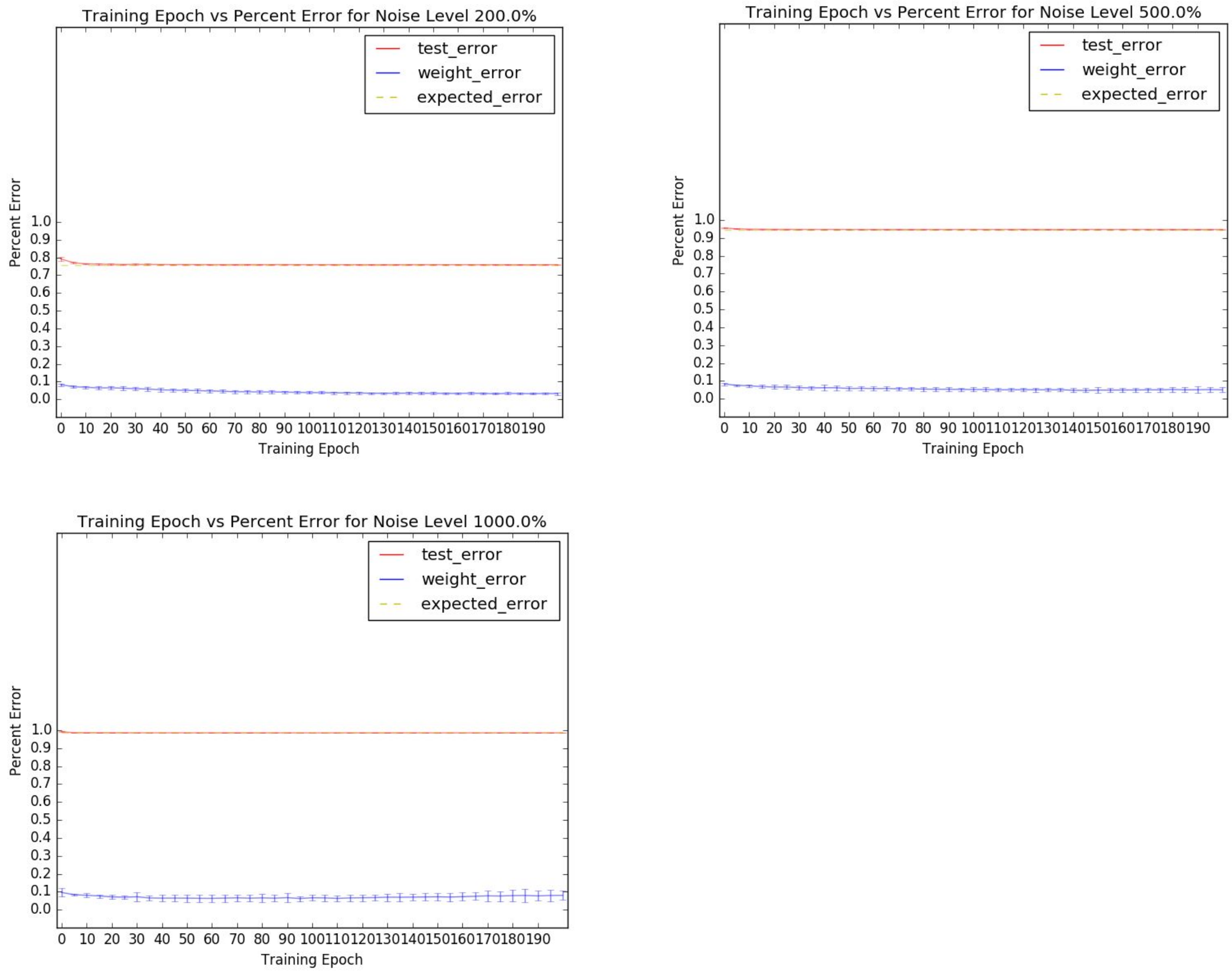

**Figure 2S.** Prediction and rule extraction accuracy of contextual regression on the simulated datasets under very high noise level.

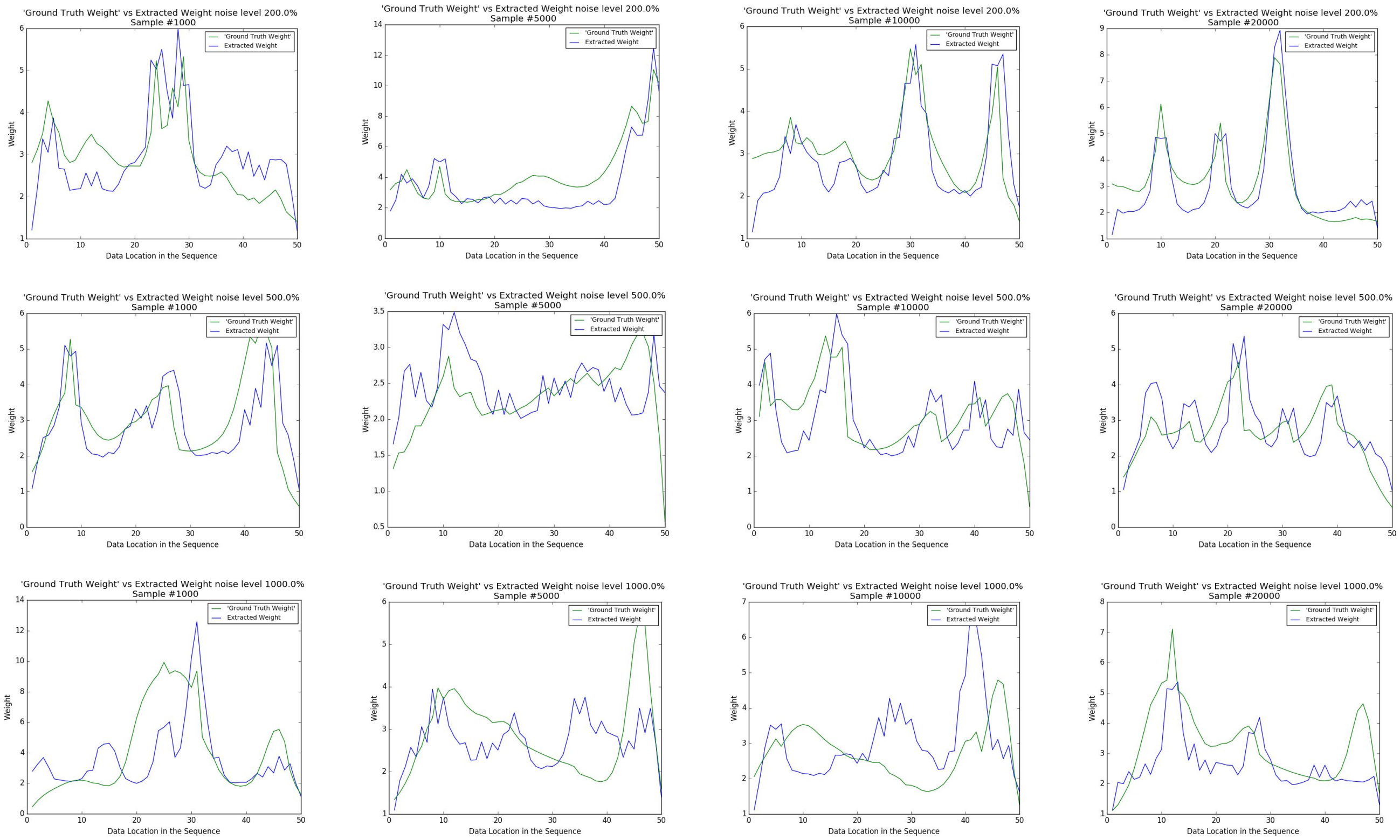

**Figure 3S.** Sample plot of “ground truth” context weight (green) vs context weight (blue) calculated by the embedding net in the contextual regression model under very high noise level.

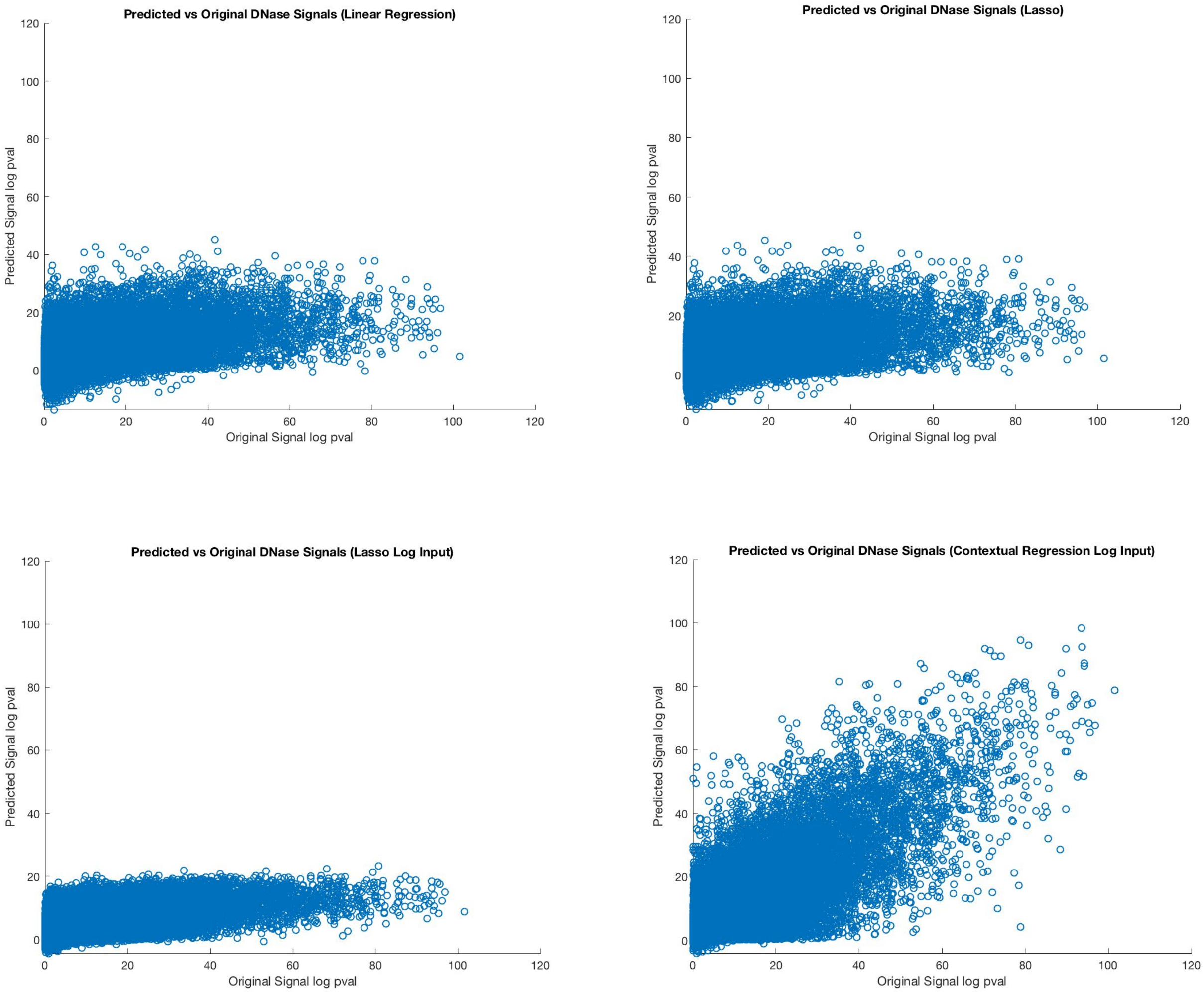

**Figure 4S.** Plot of original signal vs predicted signal by Linear Regression (top left), Lasso Regression (top right), Lasso Regression with log input (bottom left) and contextual regression with log input (bottom right) on DHSs in cell-line H1

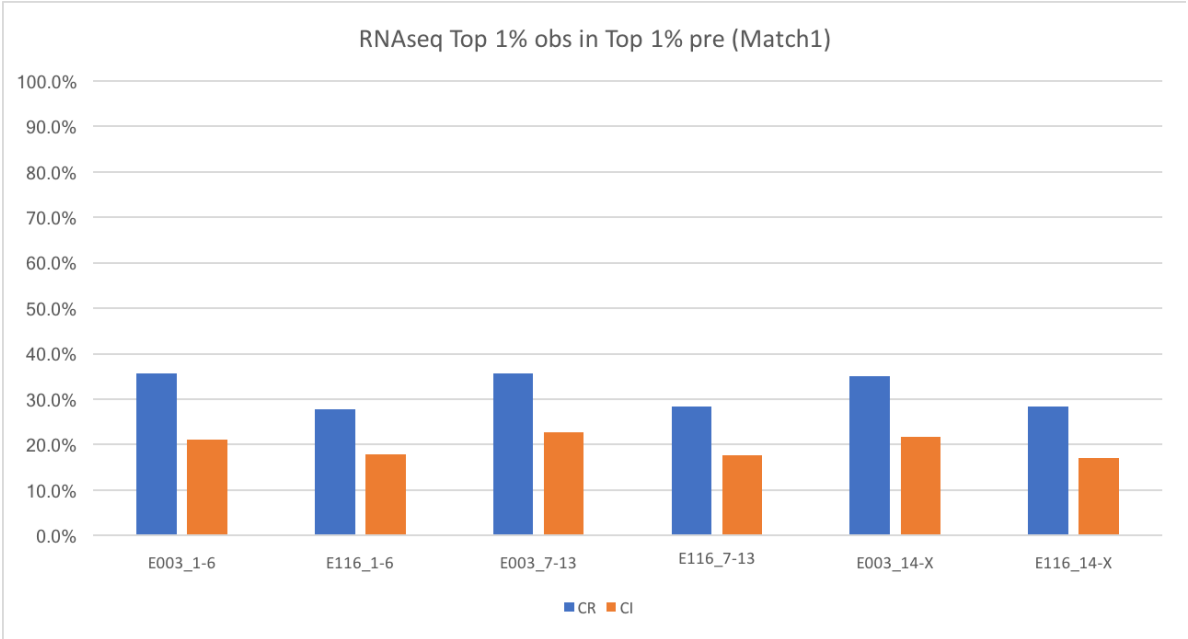

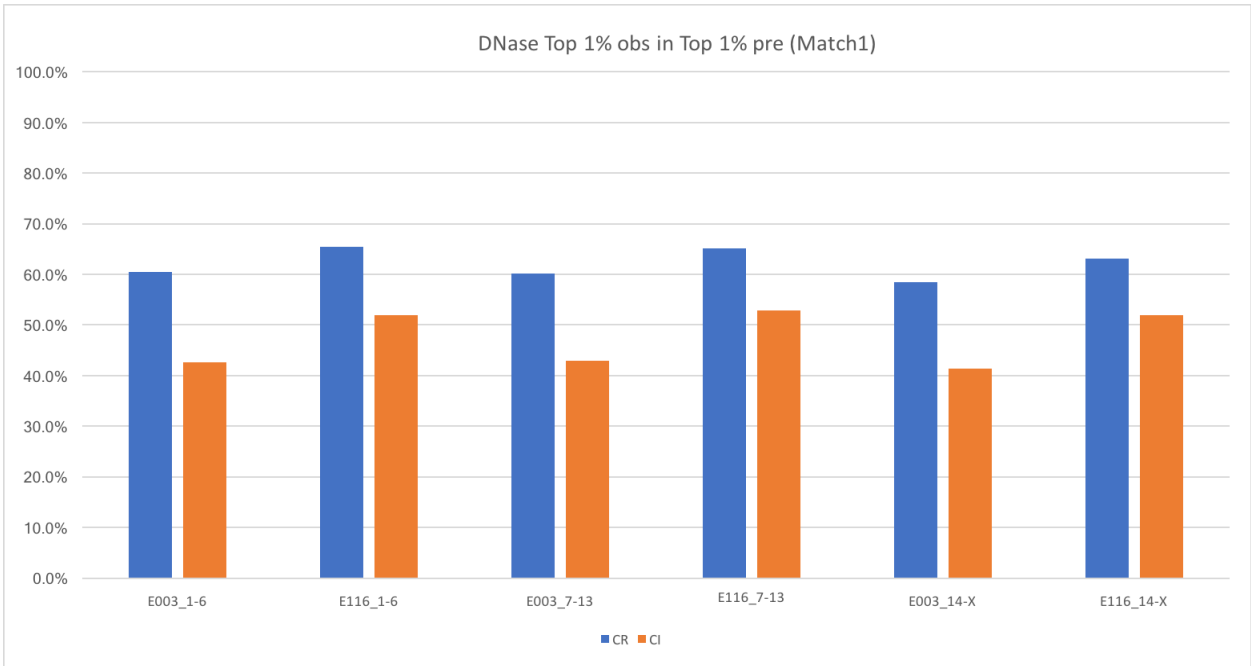

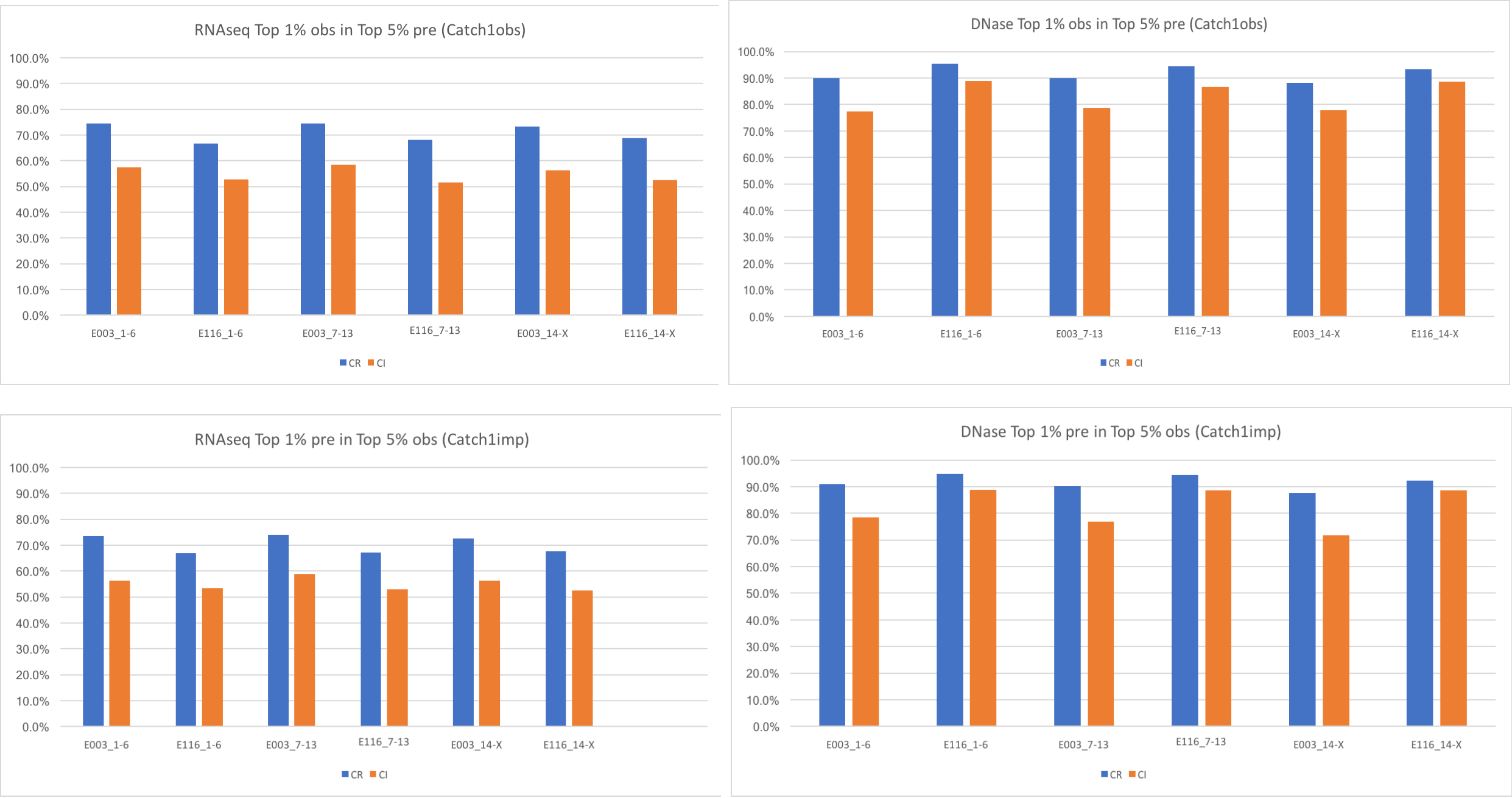

**Figure 5S.** Comparison of prediction accuracy (Match1, Catch1Obs, Catch1Imp) between contextual regression (CR) and ChromImpute (CI).

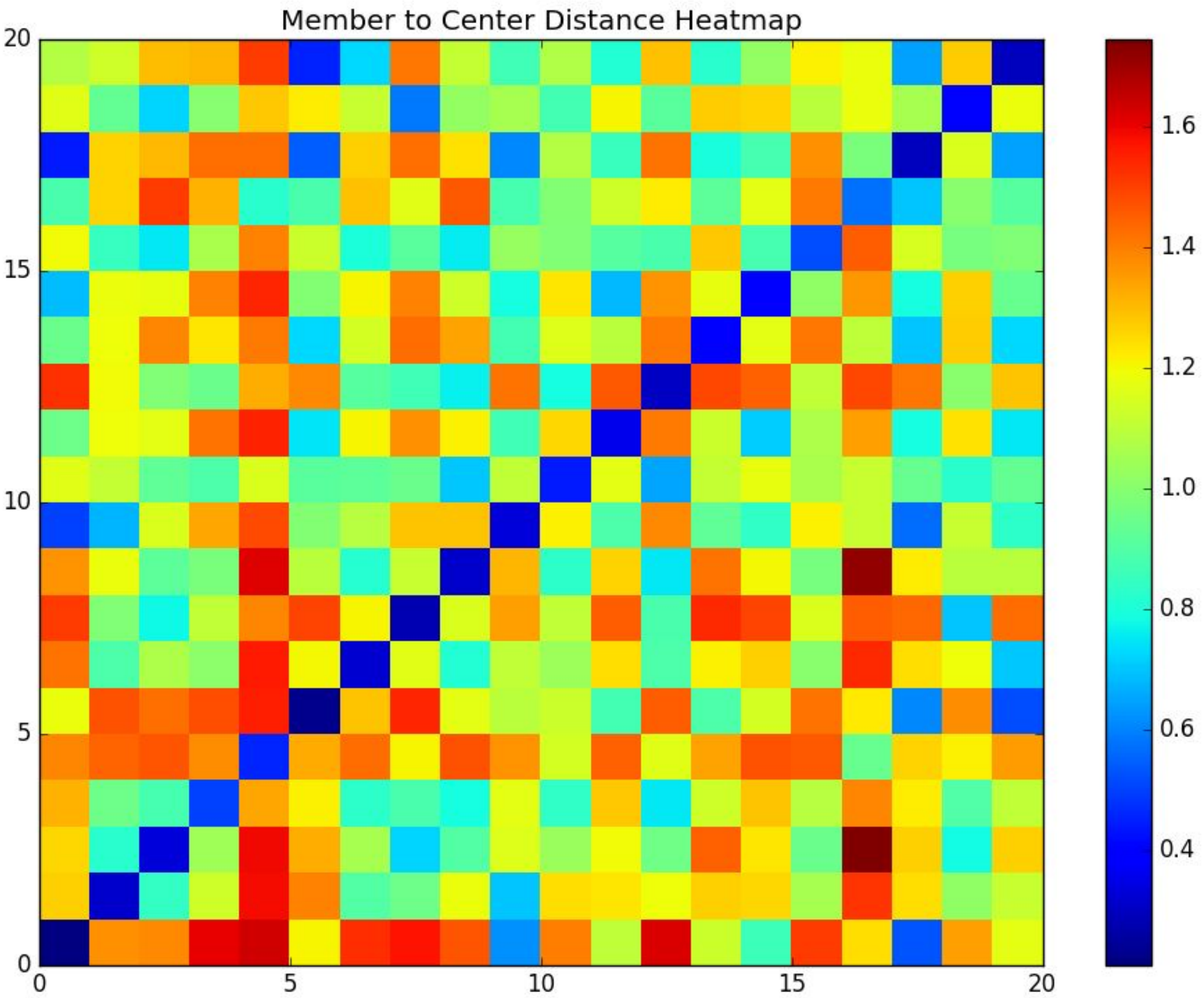

**Figure 6S.** Average in (on the main diagonal) and across (off the main diagonal) cluster distance from each cluster center

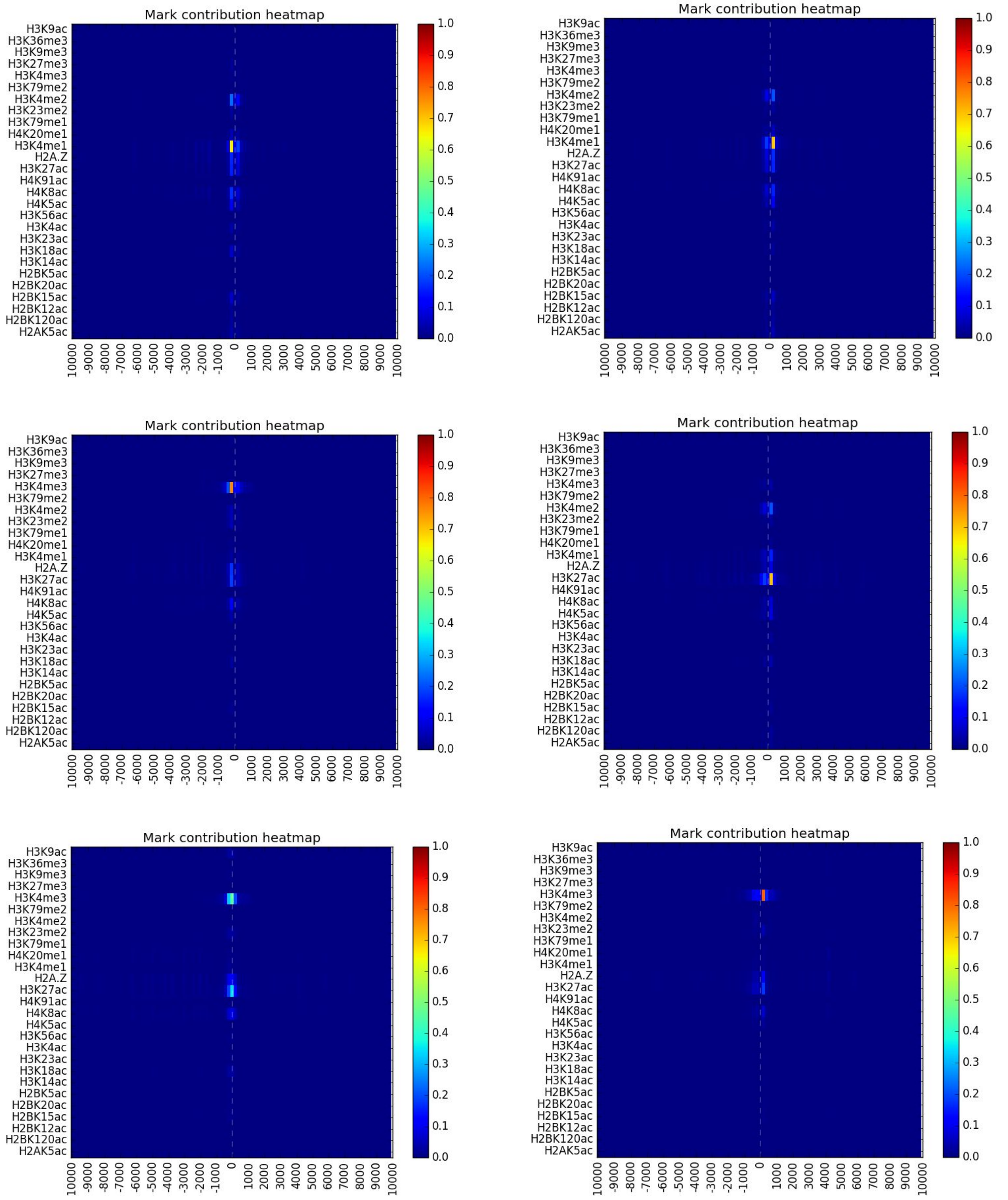

Mark contribution heatmap
H3K9ac
H3K36me3
H3K9me3
H3K27me3
H3K4me3
H3K79me2
H3K4me2
H3K23me2
H3K79me1
H4K20me1
H3K4me1
H2A.Z
H3K27ac
H4K91ac
H4K8ac
H4K5ac
H3K56ac
H3K4ac
H3K23ac
H3K18ac
H3K14ac
H2BK5ac
H2BK20ac
H2BK15ac
H2BK12ac
H2BK120ac
H2AK5ac
10000 -9000 -8000 -7000 -6000 -5000 -4000 -3000 -2000 -1000 0 1000 2000 3000 4000 5000 6000 7000 8000 9000 10000
0.0 0.1 0.2 0.3 0.4 0.5 0.6 0.7 0.8 0.9 1.0

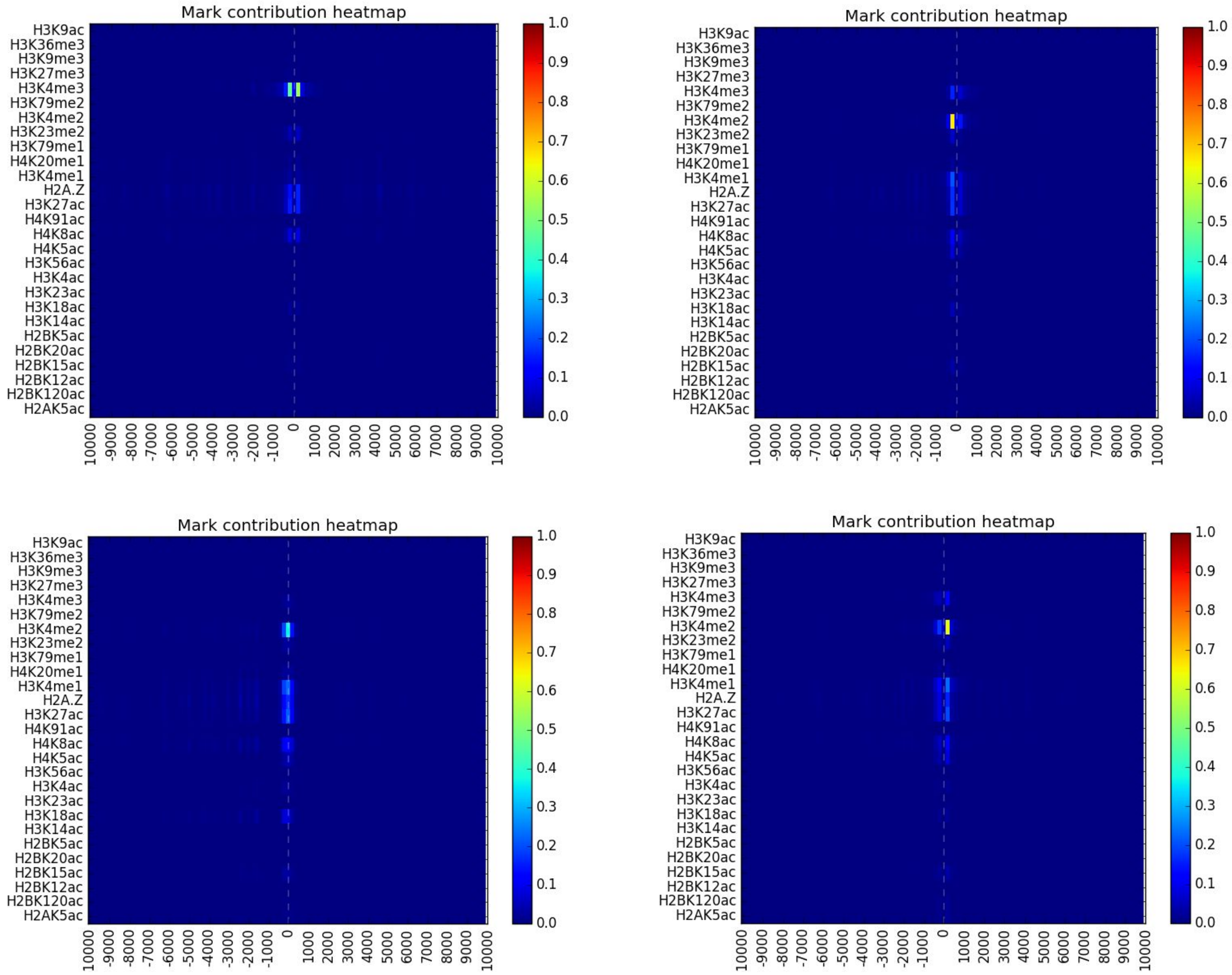
Mark contribution heatmap
Mark contribution heatmap
Mark contribution heatmap
Mark contribution heatmap
H3K9ac
H3K36me3
H3K9me3
H3K27me3
H3K4me3
H3K79me2
H3K4me2
H3K23me2
H3K79me1
H4K20me1
H3K4me1
H2A.Z
H3K27ac
H4K91ac
H4K8ac
H4K5ac
H3K56ac
H3K4ac
H3K23ac
H3K18ac
H3K14ac
H2BK5ac
H2BK20ac
H2BK15ac
H2BK12ac
H2BK120ac
H2AK5ac
10000 -9000 -8000 -7000 -6000 -5000 -4000 -3000 -2000 -1000 0 1000 2000 3000 4000 5000 6000 7000 8000 9000 10000
0.0 0.1 0.2 0.3 0.4 0.5 0.6 0.7 0.8 0.9 1.0

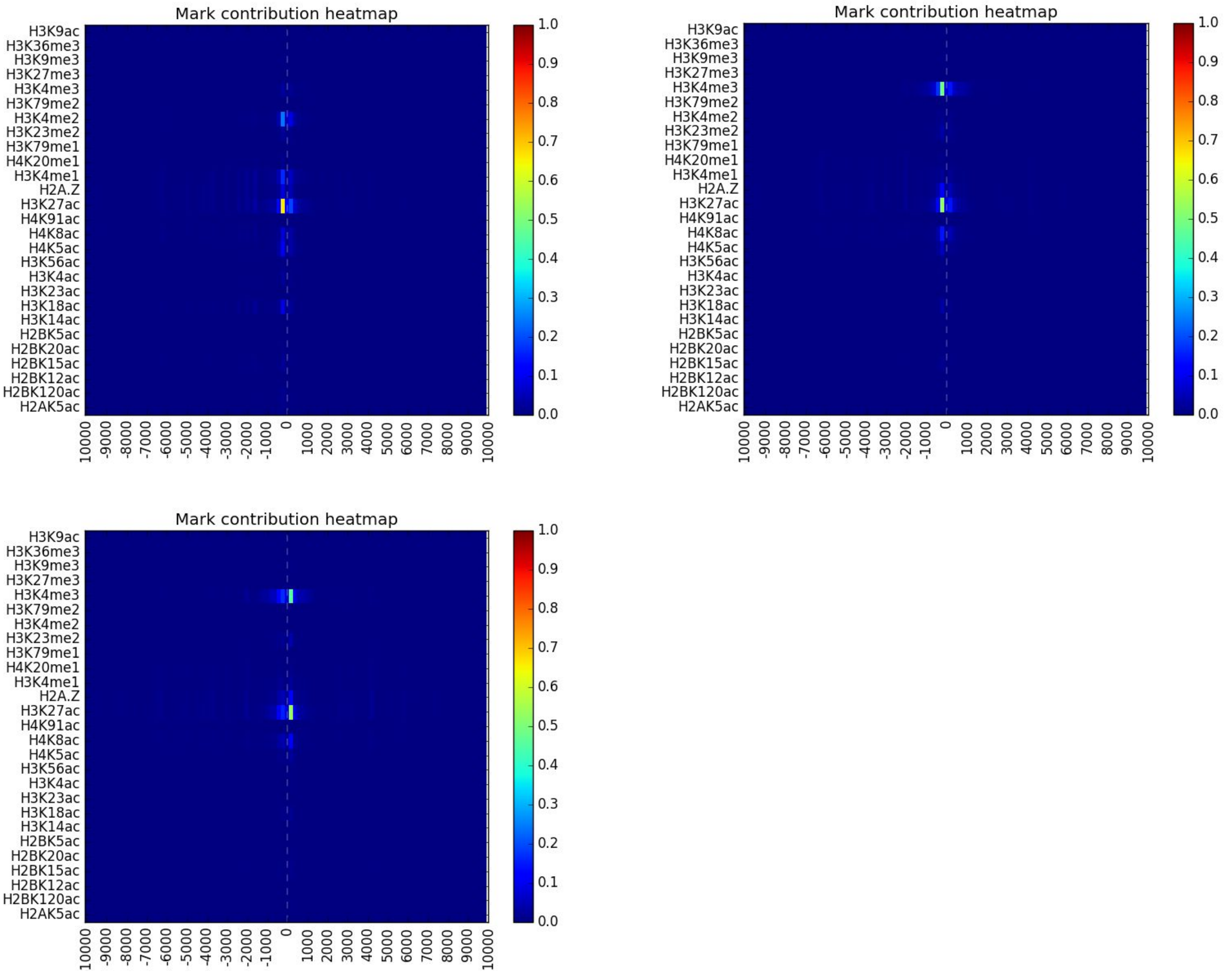

**Figure 7S.** Examples of the Type 1 (central dominant histone marks) cluster mark contribution patterns

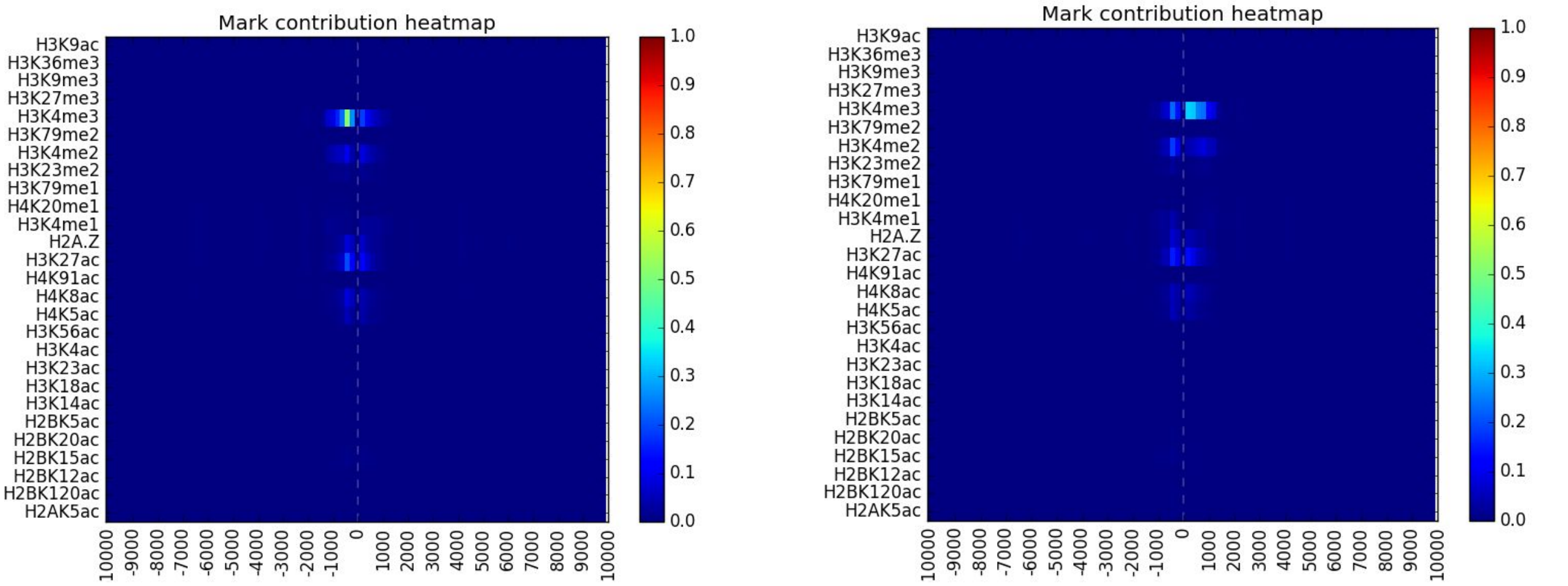

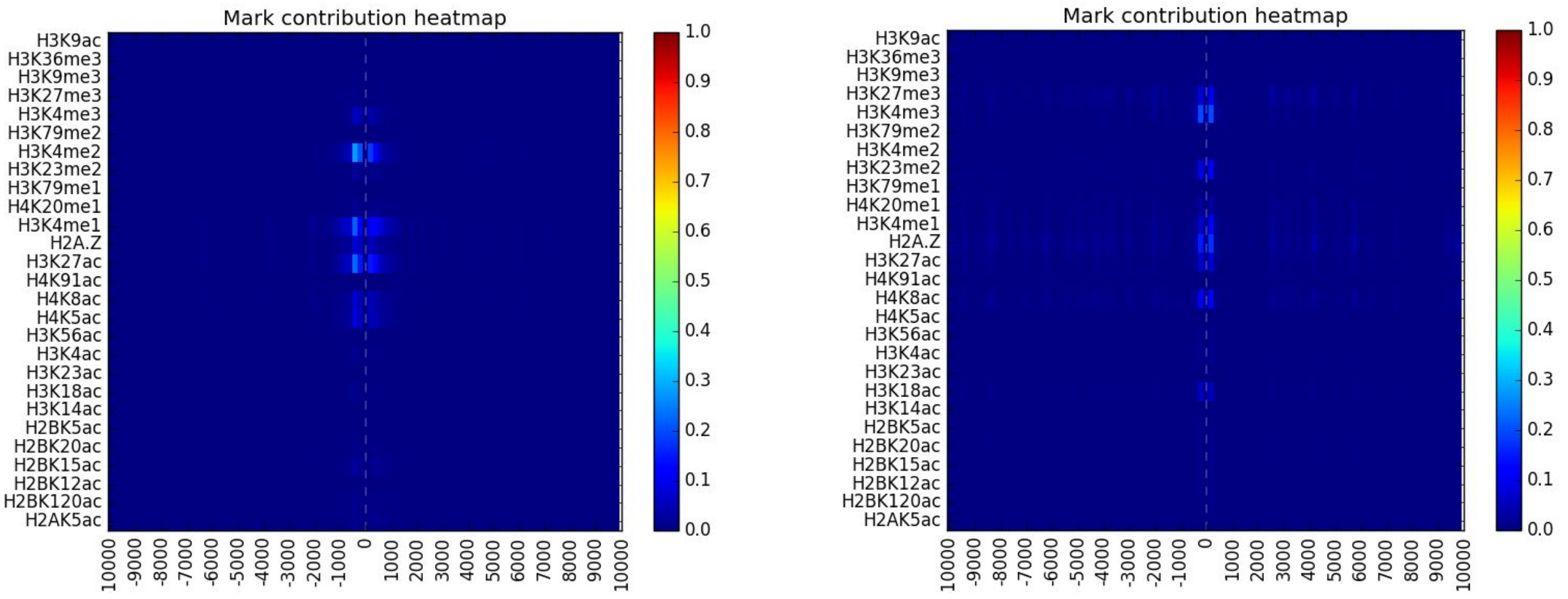

**Figure 8S.** Type 2 (spread histone marks) cluster mark contribution patterns

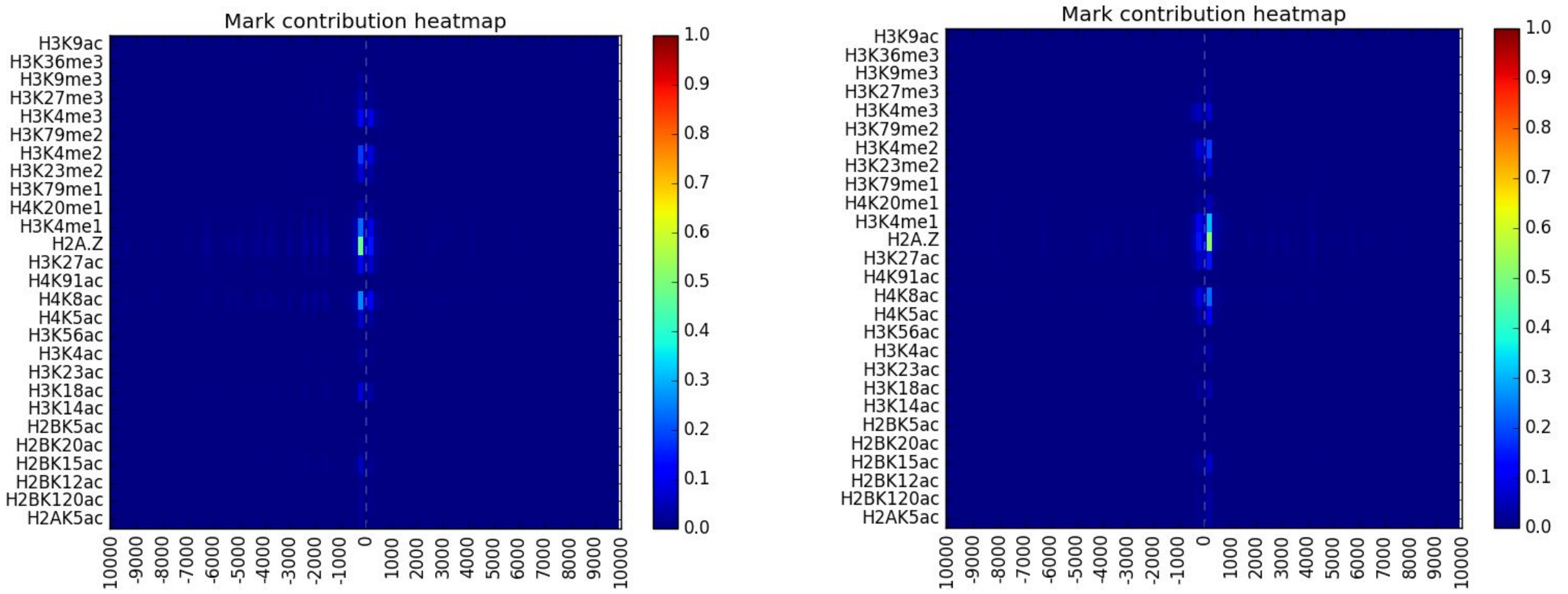

**Figure 9S.** Type 3 cluster mark (central dominant H2A.Z) contribution patterns

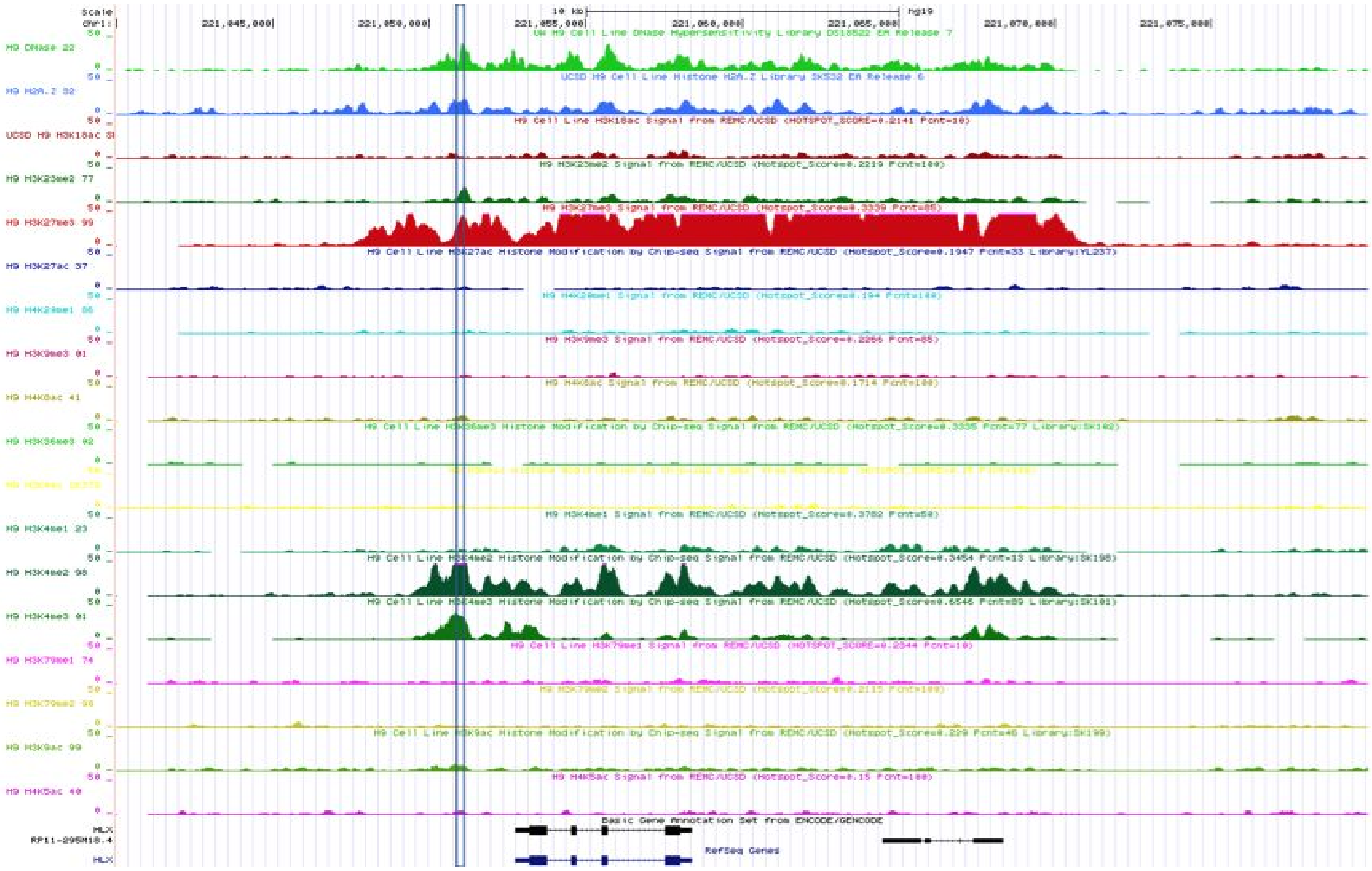

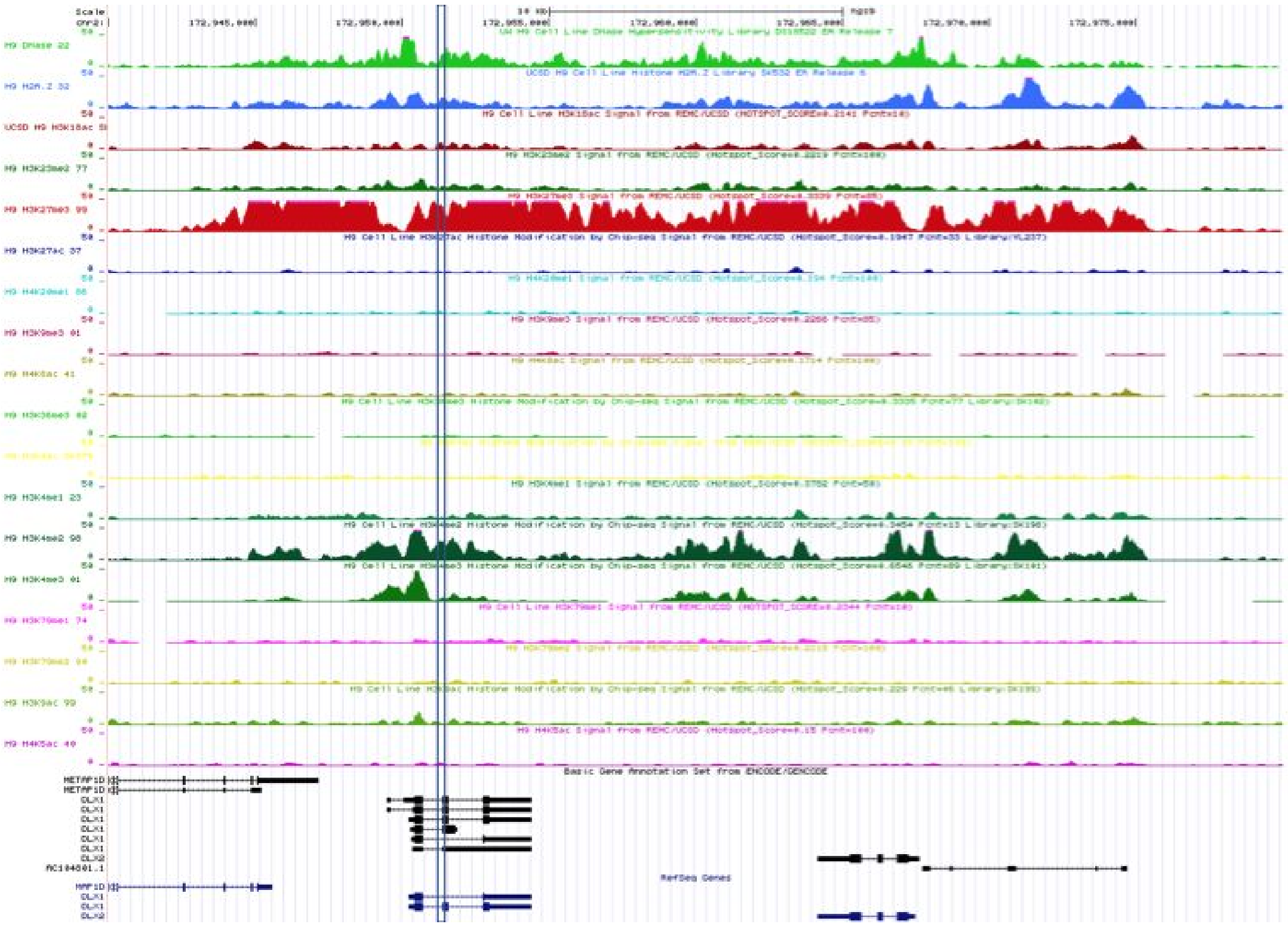

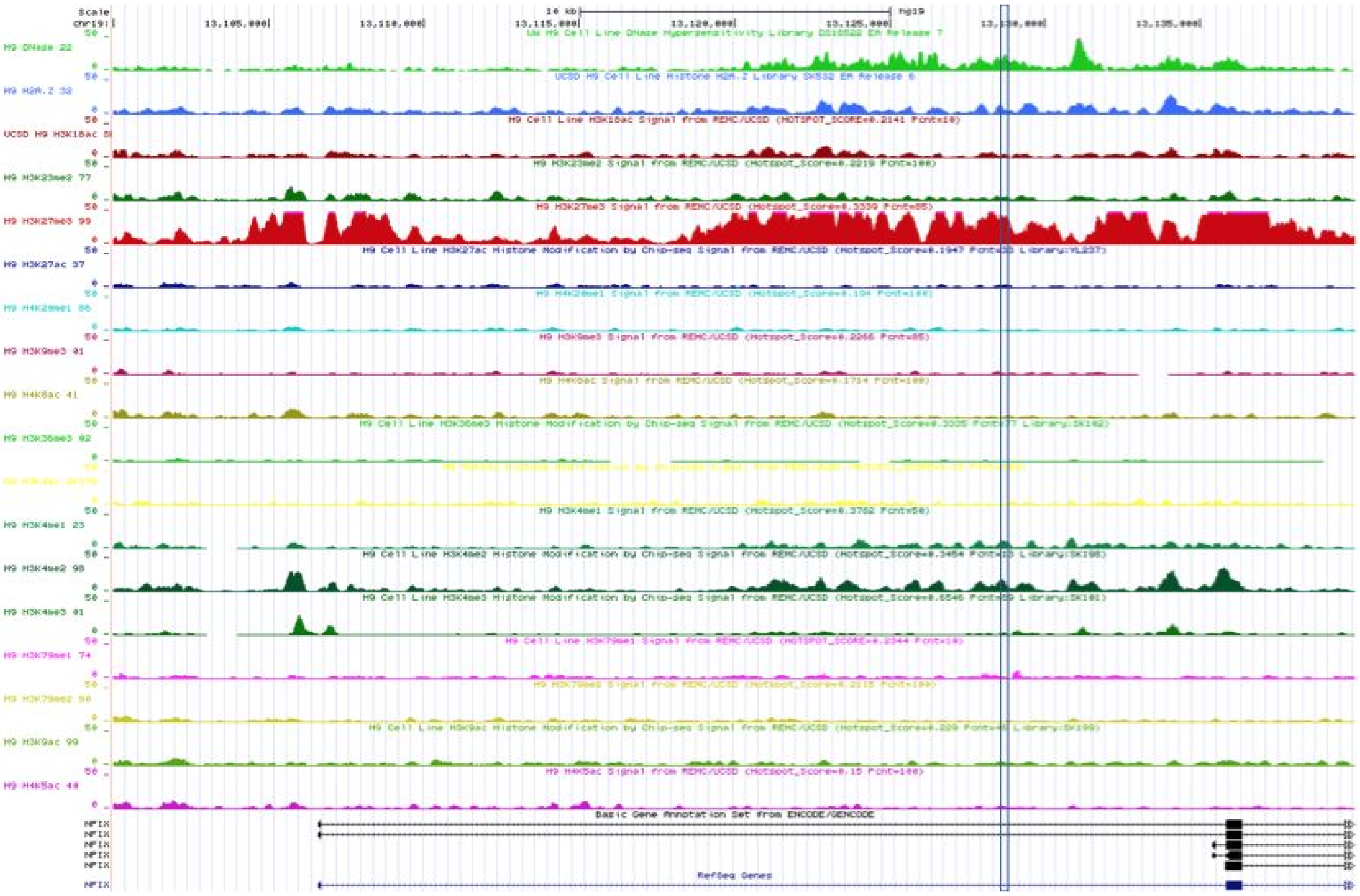
Basic Gene Annotation Set from ENCODE/GENCODE
NFIX
RefSeq Genes

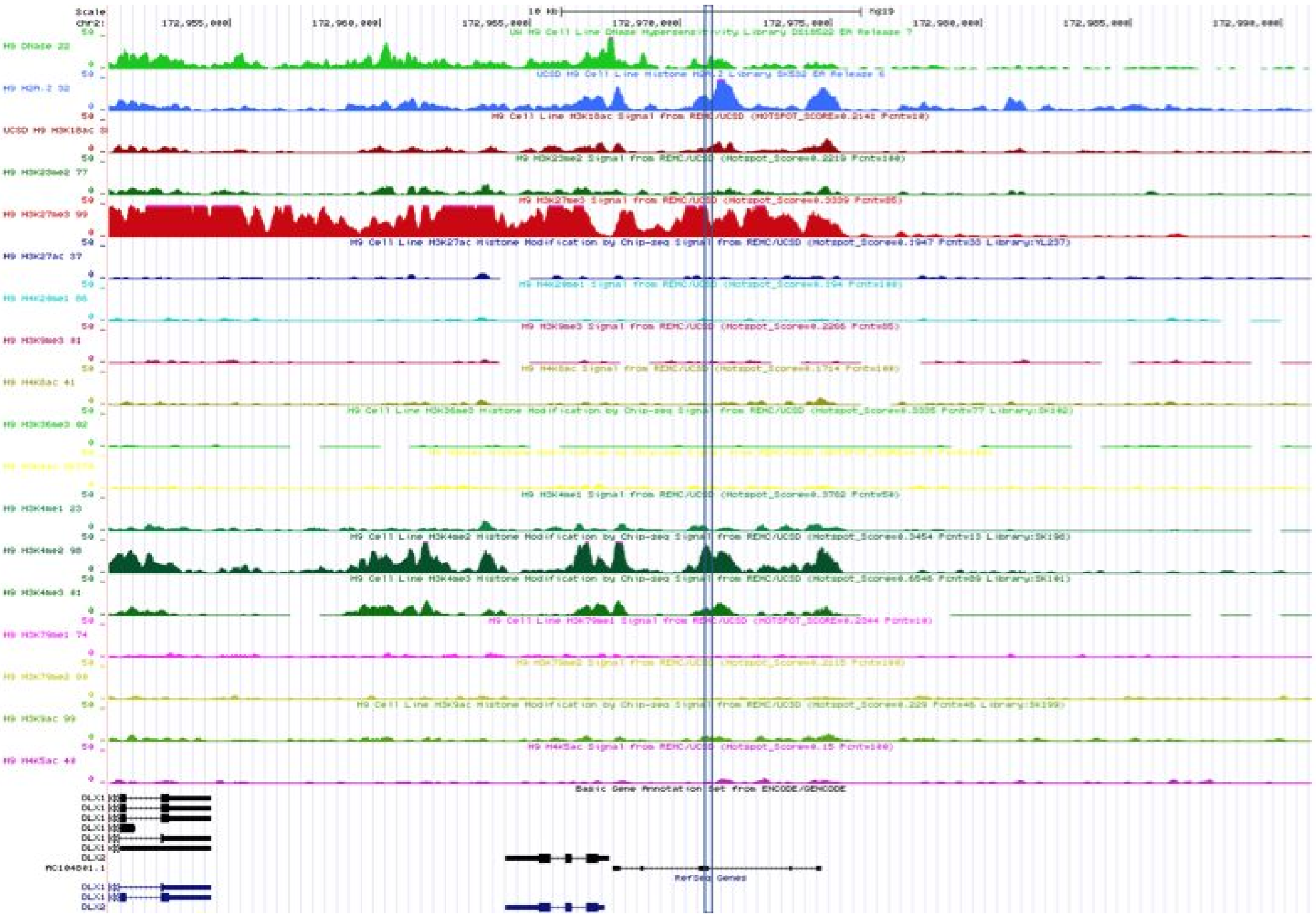
Basic Gene Annotation Set from ENCODE/GENCODE
DLX1
DLX2
RefSeq Genes

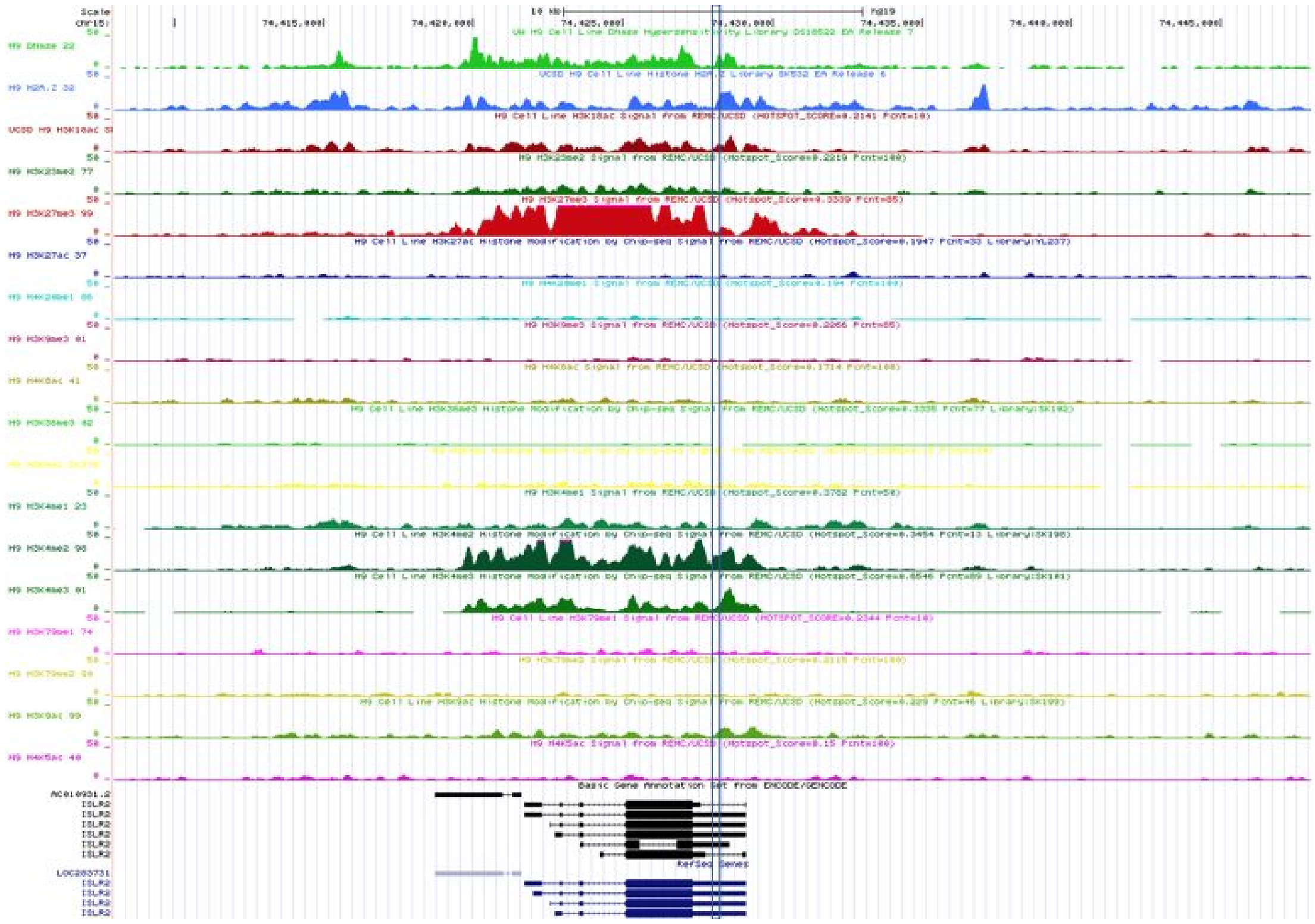

**Figure 10S.** Example genome browser views of cluster 4 in H1 (H3K27me3 cluster), the blue square is the location of the peak in the cluster detected by our method. Most of these regions include a transcription start site (TSS) in their neighborhood.

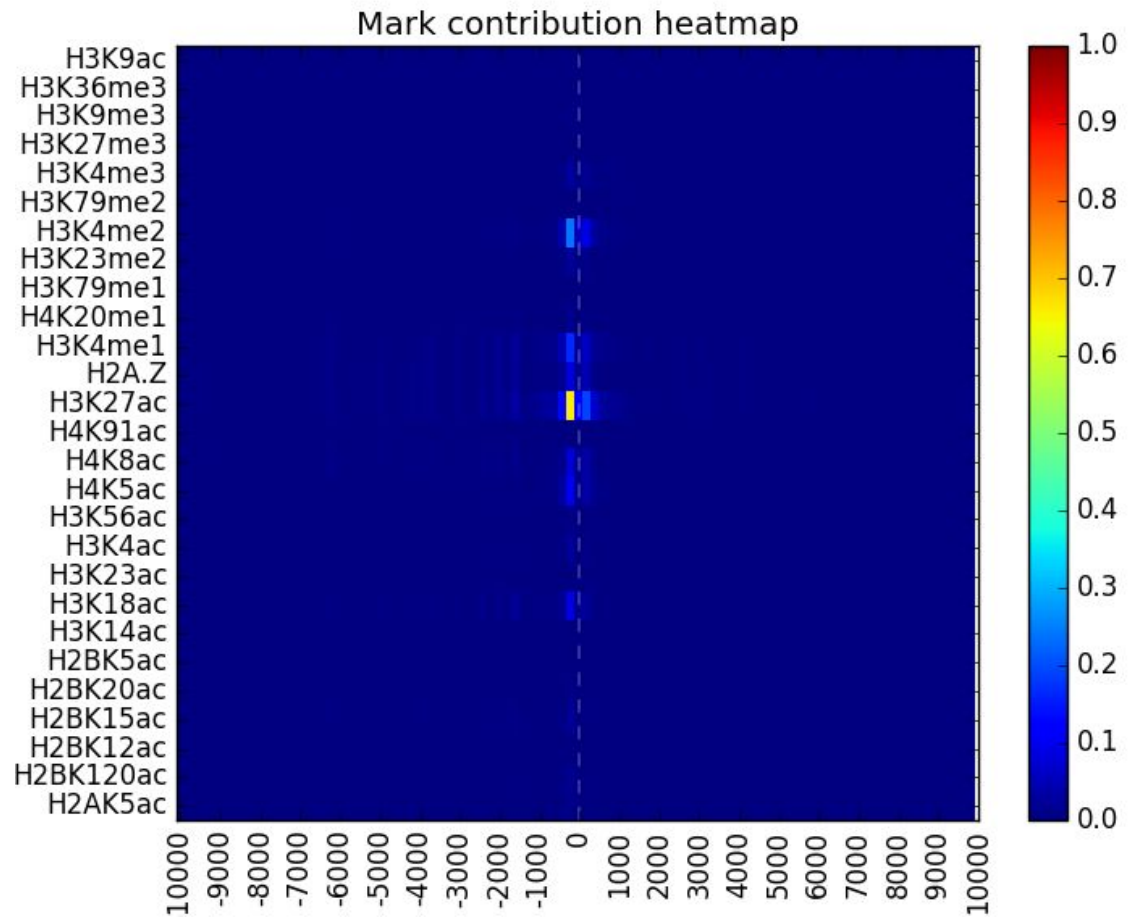

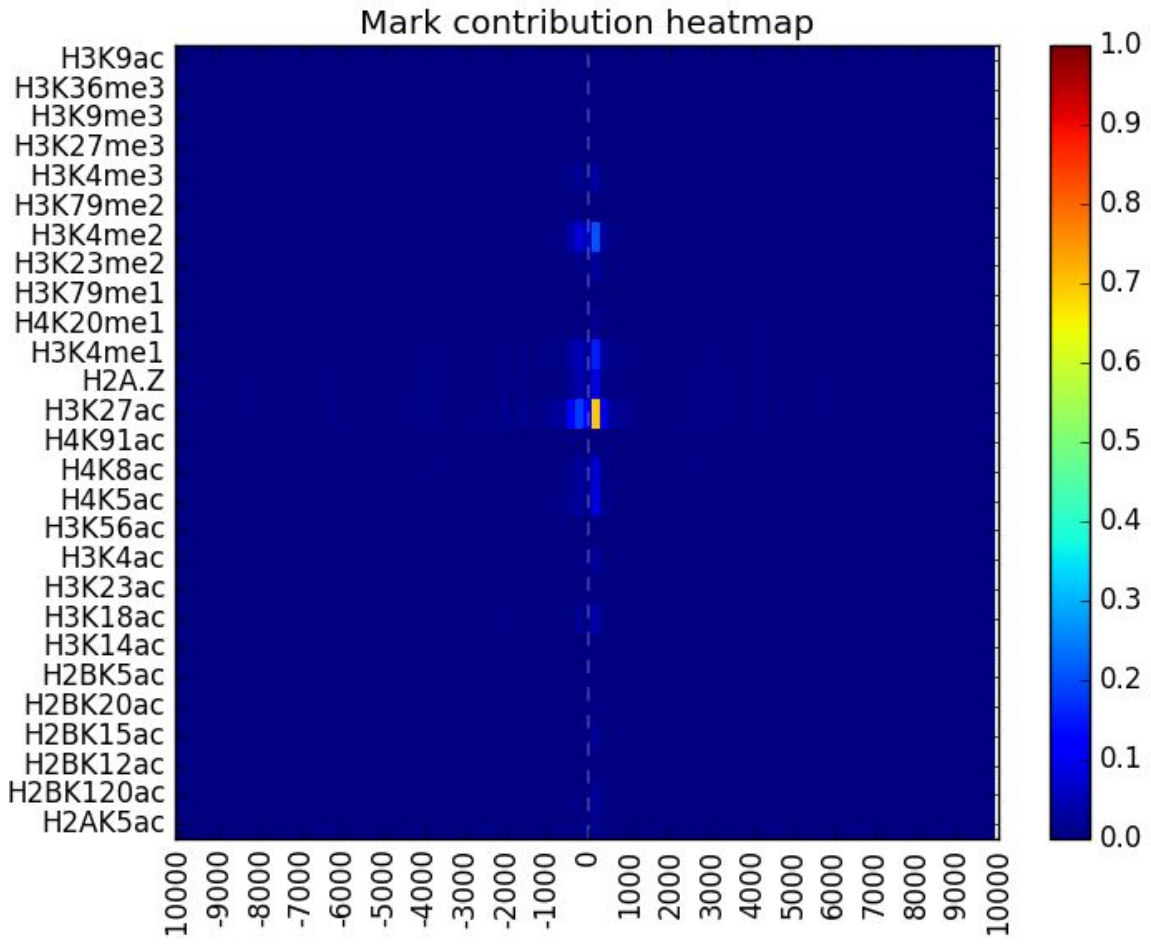

**Figure 11S.** Mark contribution patterns of cluster 1 and 6 in H1, H3K27ac dominant clusters

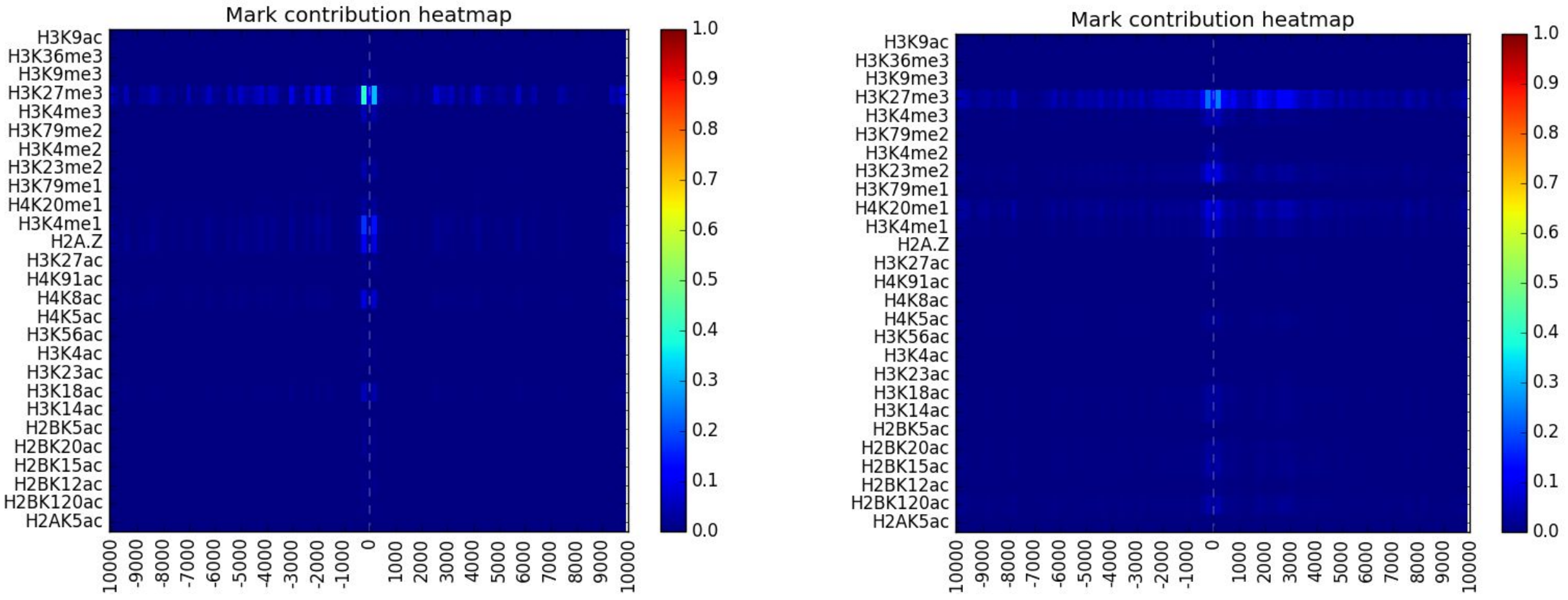

**Figure 12S.** Mark contribution pattern of cluster 4 in H1 (left) and H9 (right), both are H3K27me3 patterns

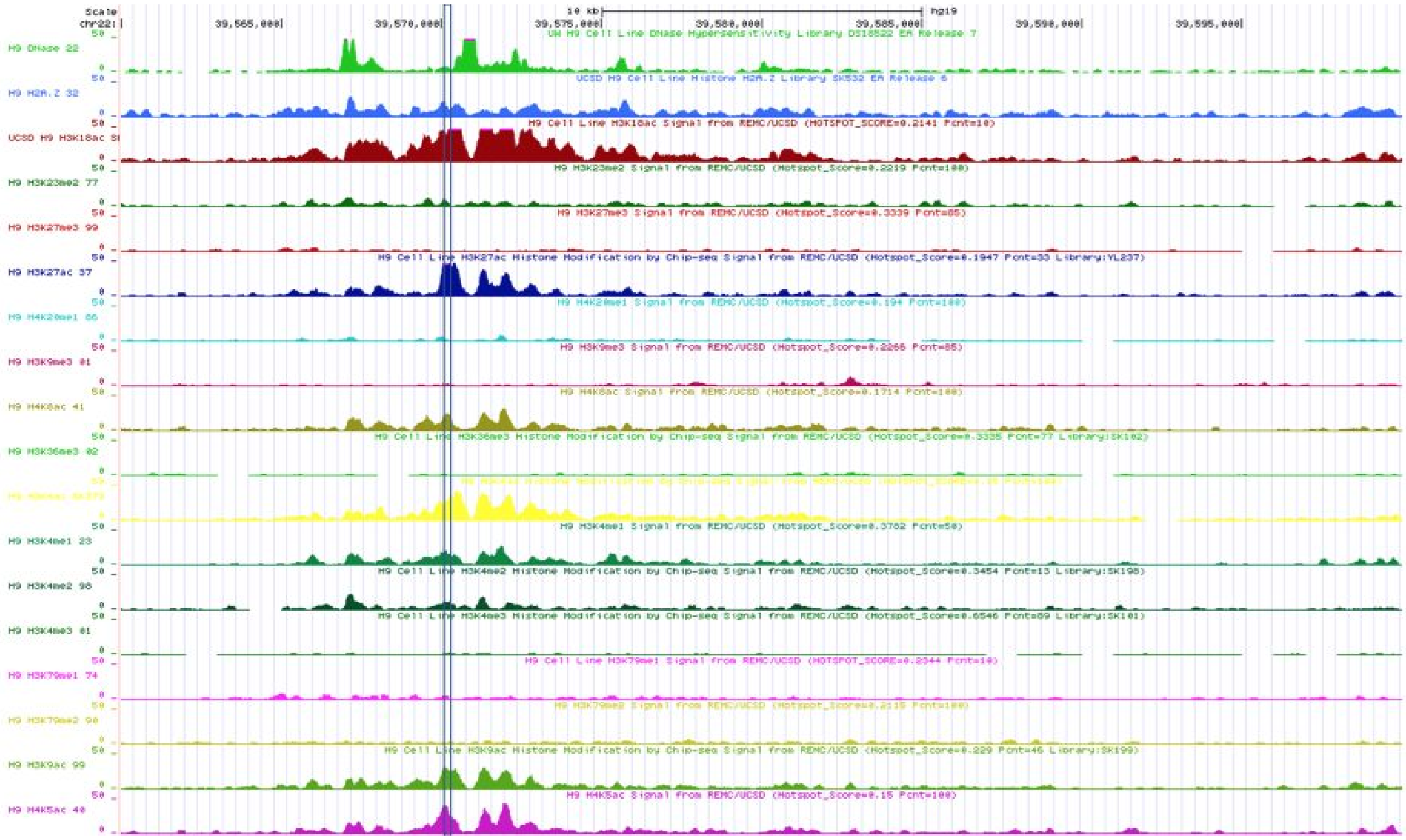

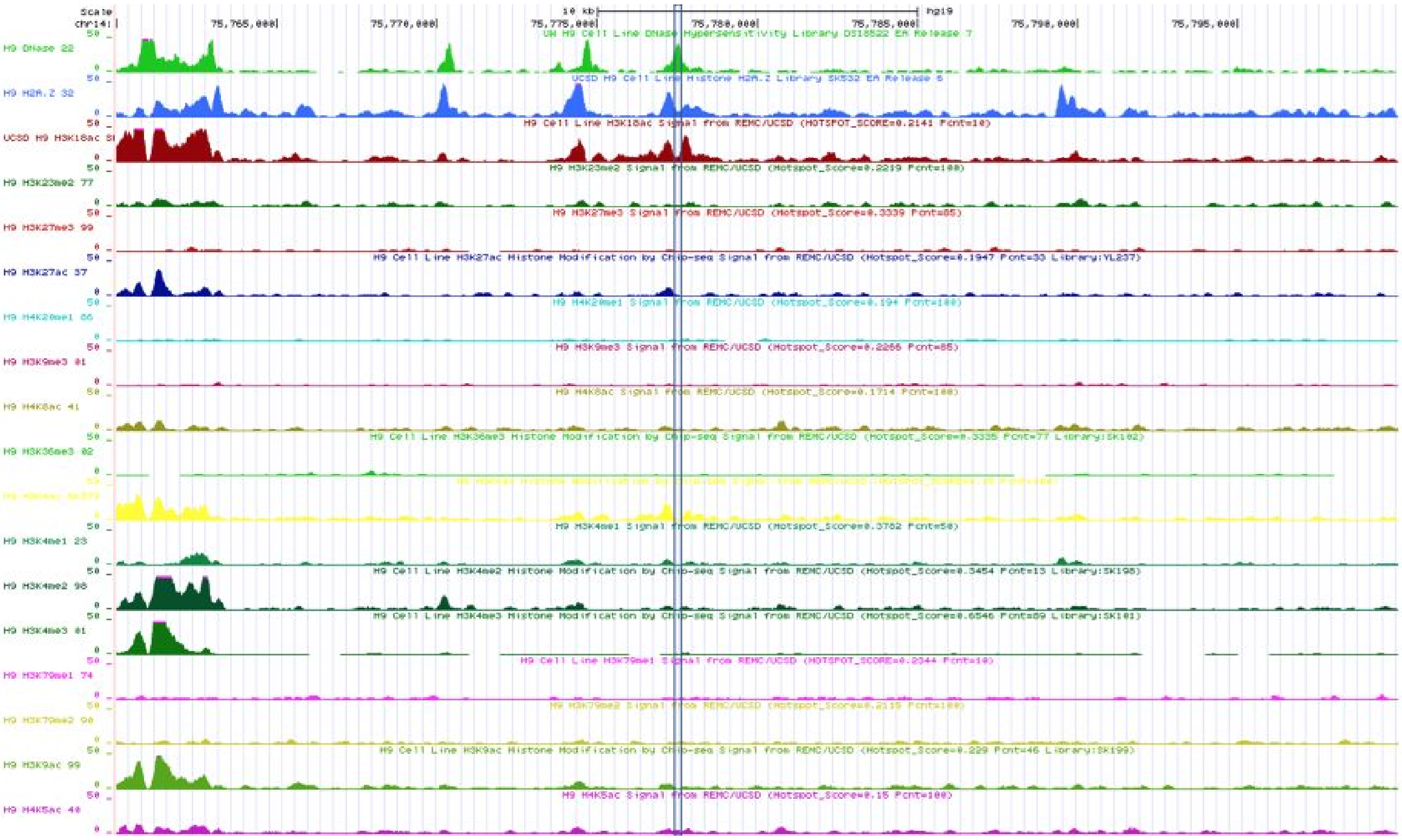

**Figure 13S.** Example genome browser views of the cluster 15 (H3K18ac dominant cluster)

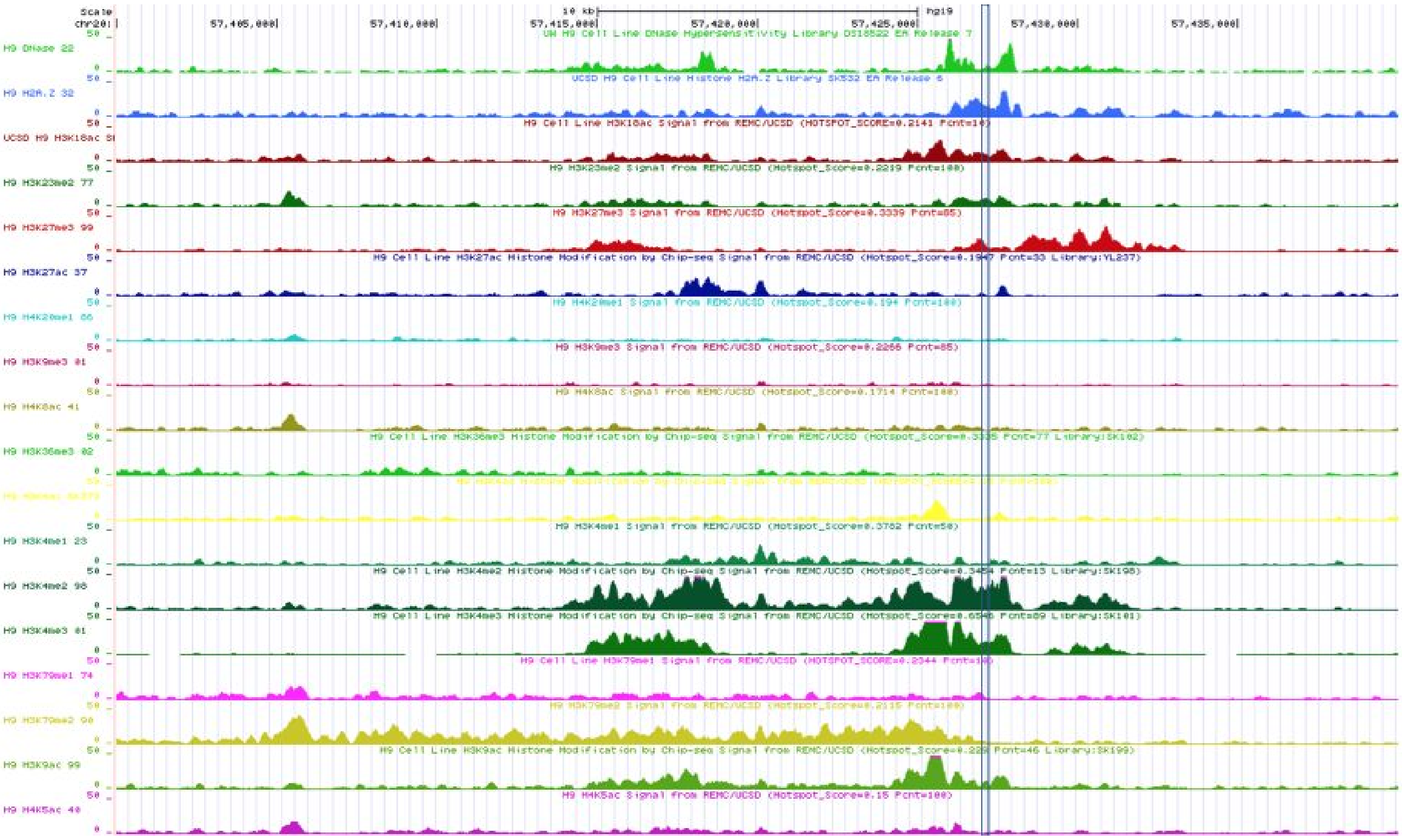

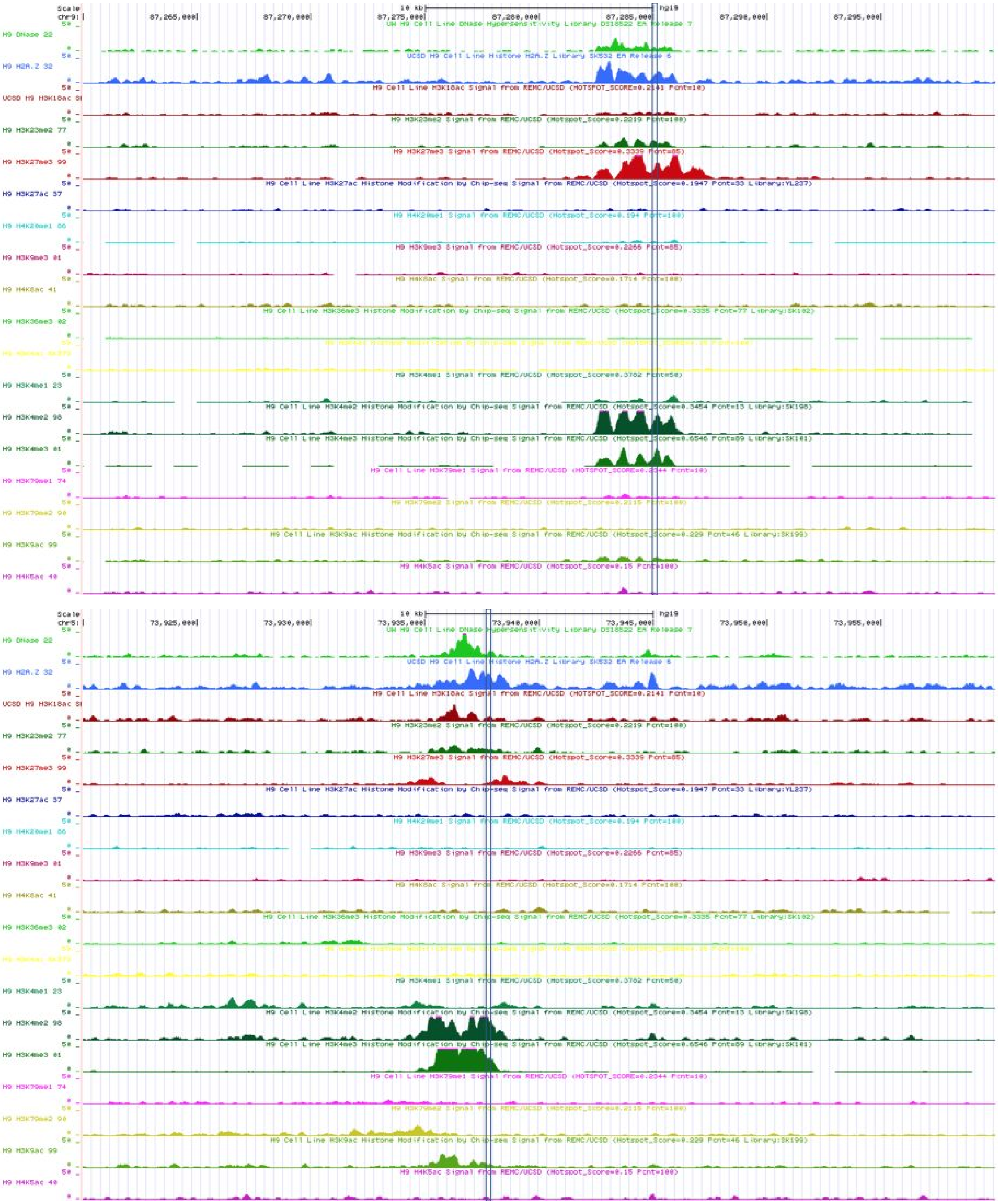

**Figure 14S.** Example Genome browser views of the cluster 5 (H3K23me2 pattern)

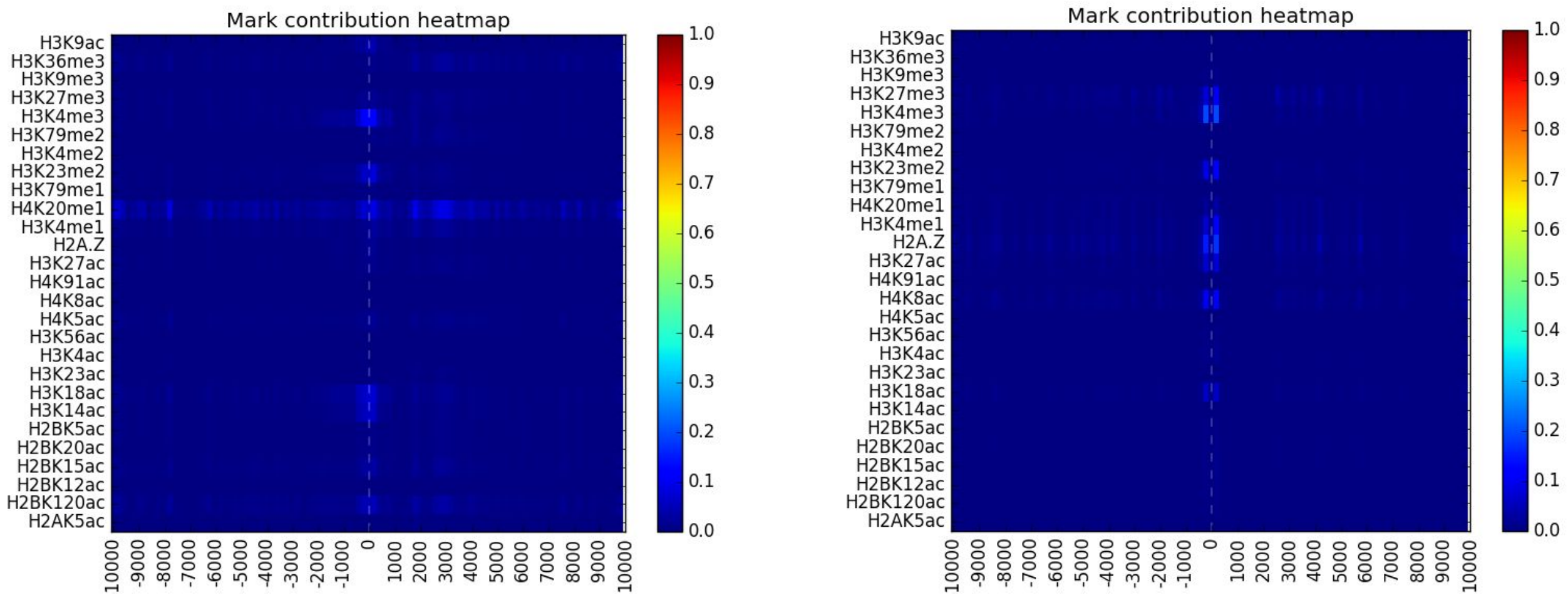

**Figure 15S.** Mark contribution patterns of cluster 16 in H1 and cluster 0 in H9 (spread type)

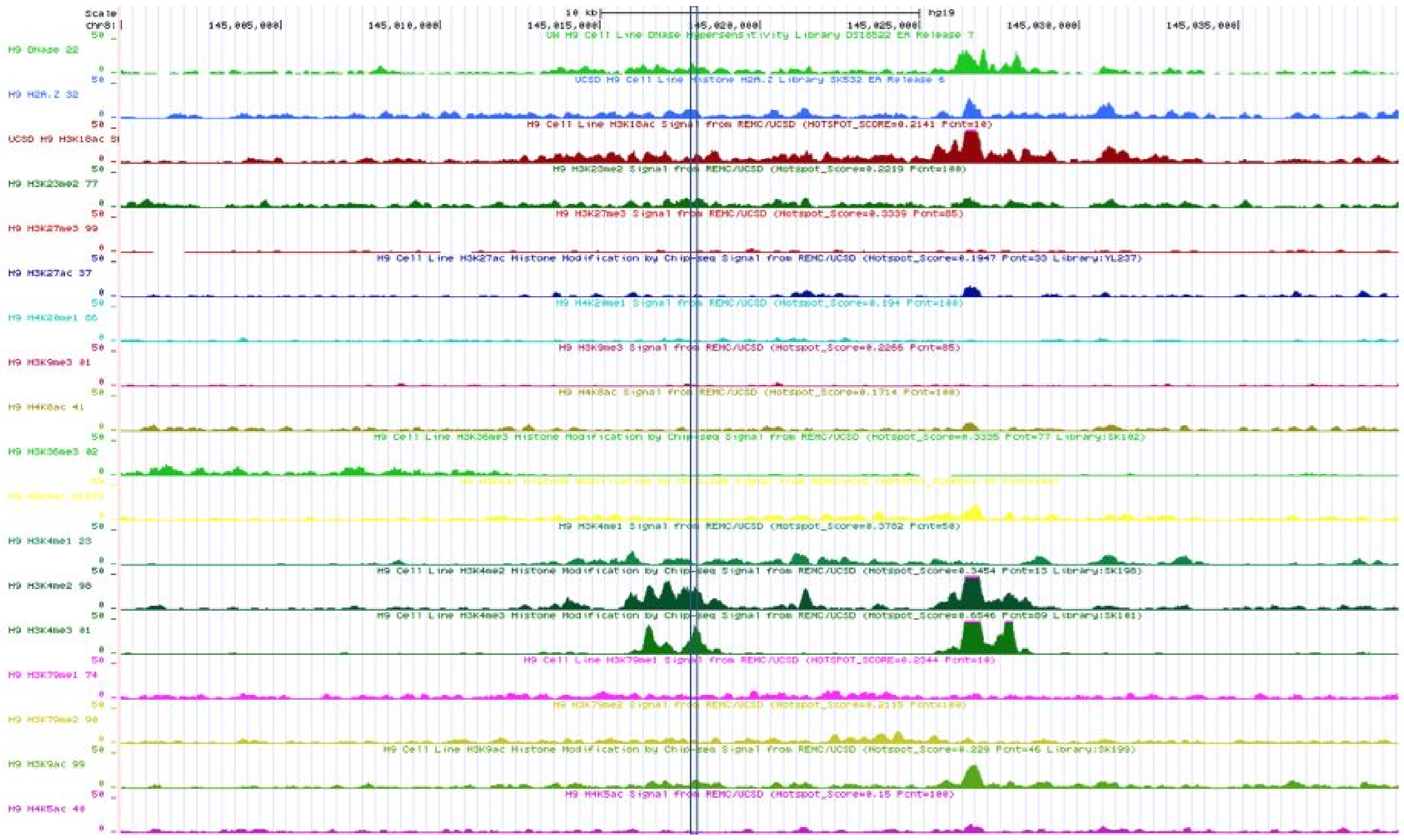

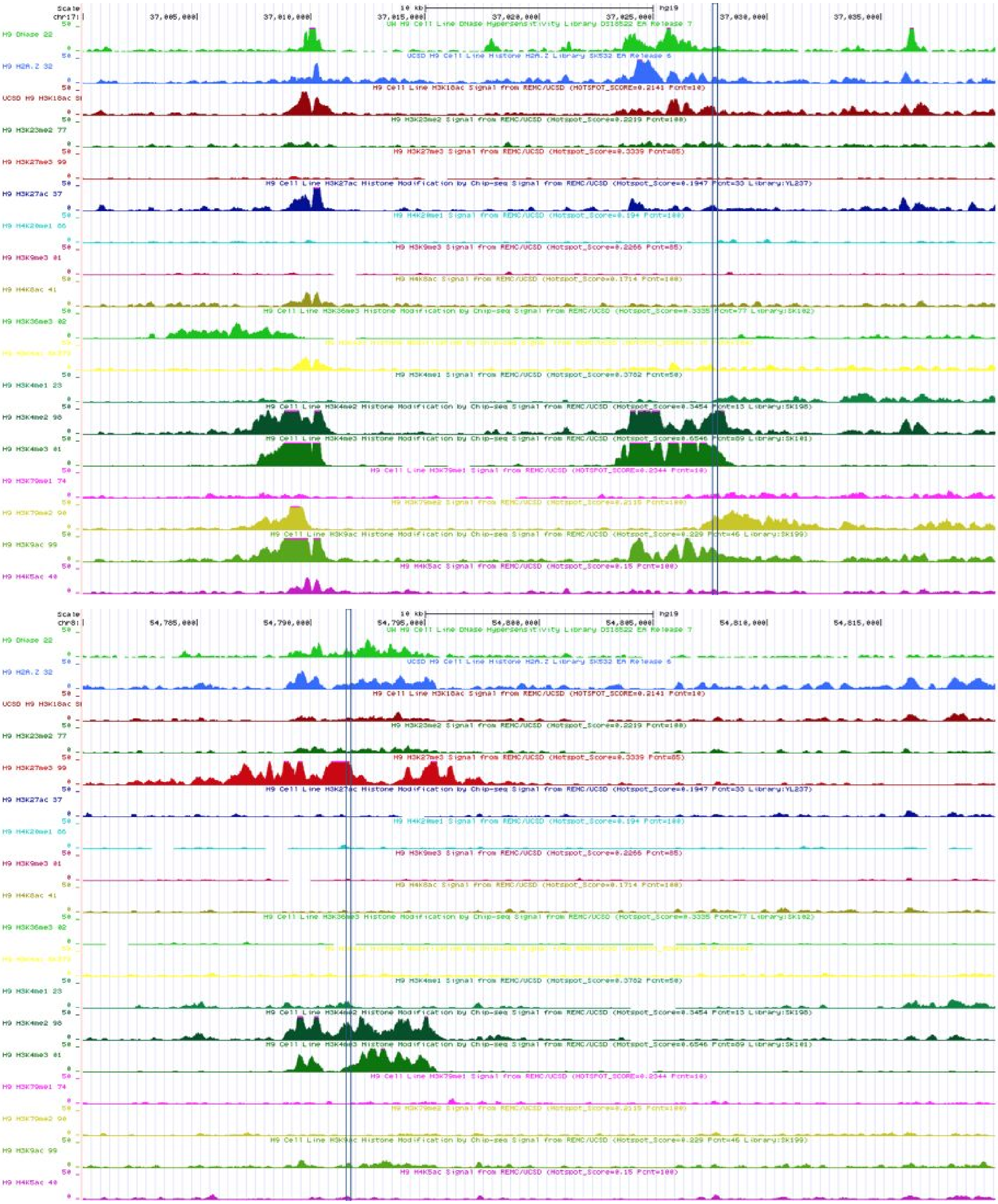

**Figure 16S.** Example genome browser views of the cluster 0 in H9 (spread type)

## Supplementary Tables:

| Testing set performance | noise level ±0% | noise level ±10% | noise level ±20% | noise level ±40% | noise level ±80% |
| --- | --- | --- | --- | --- | --- |
| Average relative prediction error (expected error) | 0.9% (0.0%) | 6.0% (5.7%) | 11.7% (11.5%) | 22.7% (22.5%) | 42.0% (41.9%) |
| Average RMSCD | 0.024 | 0.022 | 0.022 | 0.026 | 0.026 |

**Table 1S.** Prediction and rule extraction accuracy of contextual regression on a simulated dataset by the numbers

| Testing set performance | noise level ±200% | noise level ±500% | noise level ±1000% |
| --- | --- | --- | --- |
| Average relative prediction error (expected error) | 75.9% (75.6%) | 94.7% (94.5%) | 98.5% (98.5%) |
| Average RMSCD | 0.032 | 0.050 | 0.080 |

**Table 2S.** Prediction and rule extraction accuracy of contextual regression on a simulated dataset with very high noise levels

| | Pearson Correlation | Match 1 | Catch 1 Obs | Catch 1 Imp |
| --- | --- | --- | --- | --- |
| Training Set | 0.883 | 68.3% | 93.9% | 97.0% |
| Testing Set | 0.810 | 60.9% | 89.3% | 91.0% |

**Table 3S.** Prediction accuracy of the model used for histone mark pattern discovery

| H1-Cluster 4 | TX start | TX end | CDS start | CDS end |
| --- | --- | --- | --- | --- |
| in 1 kbp | 38.4% | 16.0% | 33.1% | 21.7% |
| in 3 kbp | 64.4% | 44.0% | 61.2% | 47.5% |
| in 5 kbp | 78.3% | 60.3% | 76.5% | 64.2% |
| in 7 kbp | 84.6% | 69.4% | 83.5% | 70.7% |
| in 10 kbp | 92.3% | 77.8% | 91.3% | 77.9% |

**Table 4S.** Percentage of neighborhoods of regions in H1-cluster 4 (H3K27me3 pattern) that contains TX (transcription) and CDS (coding sequence) start and end sites

| H9-Cluster 19 | Frequency Diff | p-value | Frequency in Cluster | Background Frequency |
|---|---|---|---|---|
| PO5F1 | 15.7% | 2.17e-119 | 83.0% | 67.3% |
| SOX2 | 14.9% | 2.29e-103 | 79.4% | 64.5% |
| PO2F1/2 | 14.3% | 1.56e-86 | 63.2% | 49.0% |
| NFAC2 | 14.1% | 9.35e-85 | 66.9% | 52.8% |
| PO3F2 | 13.3% | 1.05e-76 | 67.6% | 54.3% |
| NKX31 | 13.1% | 5.92e-78 | 74.8% | 61.6% |
| PIT1 | 12.6% | 6.23e-72 | 73.4% | 60.8% |

| H9-Cluster 8 | Frequency Diff | p-value | Frequency in Cluster | Background Frequency |
|---|---|---|---|---|
| PO5F1 | 15.7% | 3.16e-119 | 83.0% | 67.3% |
| SOX2 | 14.6% | 3.21e-99 | 79.1% | 64,5% |
| PO2F1/2 | 13.1% | 2.51e-73 | 62.1% | 49.0% |
| PIT1 | 13.0% | 1.55e-76 | 73.8% | 60.8% |
| PO3F1 | 12.4% | 8.89e-66 | 63.0% | 50.7% |
| PO3F2 | 12.1% | 2.45e-63 | 66.4% | 54.3% |
| NFAC2 | 12.0% | 6.17e-62 | 64.8% | 52.8% |

**Table 5S.** Motif enrichment profile of the two H3K27ac dominant clusters in H9 (cluster 8 and 19)

| H1-Cluster 1 | Frequency Diff | p-value | Frequency in Cluster | Background Frequency |
|---|---|---|---|---|
| SOX2 | 7.6% | 1.94e-29 | 71.0% | 63.4% |
| PO5F1 | 7.0% | 4.83e-26 | 72.9% | 65.9% |
| PO2F1/2 | 6.0% | 7.25e-18 | 53.8% | 47.8% |
| SOX9 | 5.3% | 1.08e-14 | 62.5% | 57.2% |
| PO3F1 | 5.0% | 4.53e-13 | 54.6% | 49.6% |

| H1-Cluster 6 | Frequency Diff | p-value | Frequency in Cluster | Background Frequency |
|---|---|---|---|---|
| SOX2 | 7.8% | 4.59e-30 | 71.3% | 63.4% |
| PO5F1 | 7.5% | 3.07e-28 | 73.4% | 65.9% |
| PO2F1/2 | 5.1% | 8.28e-13 | 52.9% | 47.8% |
| SOX9 | 5.3% | 4.05e-14 | 62.5% | 57.2% |
| PO3F1 | 4.5% | 1.66e-10 | 54.1% | 49.6% |

**Table 6S.** Motif enrichment profile of cluster 1 and 6 in H1, H3K27ac dominant clusters

## Reference:

1. Kingma, D., Ba - arXiv preprint arXiv:1412.6980, J. & 2014. Adam: A method for stochastic optimization. *arxiv.org* (1412).
2. ENCODE Project Consortium. An integrated encyclopedia of DNA elements in the human genome. *Nature* **489,** 57–74 (2012).
3. Bannister, A. J. & Kouzarides, T. Regulation of chromatin by histone modifications. *Cell Res.* **21,** 381–395 (2011).
4. Xu, Y. *et al.* Histone H2A.Z Controls a Critical Chromatin Remodeling Step Required for DNA Double-Strand Break Repair. *Mol. Cell* **48,** 723–733 (2012).
5. Lombardi, L. *Maintenance of Open Chromatin States by Histone H3 Eviction and H2A.Z*. (2011).
6. Huang, J., Marco, E., Pinello, L. & Yuan, G.-C. Predicting chromatin organization using histone marks. *Genome Biol.* **16,** 162 (2015).
7. Gomez, N. C. *et al.* Widespread Chromatin Accessibility at Repetitive Elements Links Stem Cells with Human Cancer. *Cell Rep.* **17,** 1607–1620 (2016).
8. High-Resolution Profiling of Histone Methylations in the Human Genome. *Cell* **129,** 823–837 (2007).
9. Schwartz, S., Meshorer, E. & Ast, G. Chromatin organization marks exon-intron structure. *Nat. Struct. Mol. Biol.* **16,** 990–995 (2009).
10. Karlić, R., Chung, H.-R., Lasserre, J., Vlahoviček, K. & Vingron, M. Histone modification levels are

predictive for gene expression. *Proc. Natl. Acad. Sci. U. S. A.* **107,** 2926–2931 (2010).

11. Rintisch, C. *et al.* Natural variation of histone modification and its impact on gene expression in the rat genome. *Genome Res.* **24,** 942–953 (2014).

12. Pearson, K. *On Lines and Planes of Closest Fit to Systems of Points in Space*. (1901).

13. Atkinson, K. E. *AN INTRODUCTION TO NUMERICAL ANALYSIS, 2ND ED*. (John Wiley & Sons, 2008).